\documentclass{andp2012}
\usepackage[english]{babel}
\usepackage{setspace}
\usepackage{graphicx}
\usepackage{amssymb,amsmath}
\usepackage{subfigure}
\usepackage{graphicx}
\usepackage{dsfont}
\usepackage{xcolor}
\usepackage{rotating}

\usepackage{tabularx}
\usepackage{cuted}
\usepackage[scr=boondoxo,scrscaled=1.05]{mathalfa}

\usepackage{tikz}
\usetikzlibrary{calc}

\newcommand\tbrac[1]{\tikz[baseline=(n.base)]{
    \node(n)[inner sep=2pt]{$#1$};
    \draw[line cap=round] ($(n.west)+0.7*(n.north)$) -- ++ ($0.3*(n.north)$) -- ++ ($2*(n.east)$) -- ++ ($0.3*(n.south)$);
  }}  
\newcommand\bbrac[1]{\tikz[baseline=(n.base)]{
    \node(n)[inner sep=2pt]{$#1$};
    \draw[line cap=round] ($(n.west)+0.7*(n.south)$) -- ++ ($0.3*(n.south)$) -- ++ ($2*(n.east)$) -- ++ ($0.3*(n.north)$);
  }}  
\newcommand\lbrac[1]{\tikz[baseline=(n.base)]{
    \node(n)[inner sep=2pt]{$#1$};
    \draw[line cap=round] ($0.8*(n.west)+(n.north)$) -- ++ ($0.2*(n.west)$) -- ++ ($2*(n.south)$) -- ++ ($0.2*(n.east)$);
  }}  
\newcommand\rbrac[1]{\tikz[baseline=(n.base)]{
    \node(n)[inner sep=2pt]{$#1$};
    \draw[line cap=round] ($0.8*(n.east)+(n.north)$) -- ++ ($0.2*(n.east)$) -- ++ ($2*(n.south)$) -- ++ ($0.2*(n.west)$);
  }}
\newcommand\lrbrac[1]{\tikz[baseline=(n.base)]{
    \node(n)[inner sep=2pt]{$#1$};
    \draw[line cap=round] ($0.8*(n.west)+(n.north)$) -- ++ ($0.2*(n.west)$) -- ++ ($2*(n.south)$) -- ++ ($0.2*(n.east)$);
    \draw[line cap=round] ($0.8*(n.east)+(n.north)$) -- ++ ($0.2*(n.east)$) -- ++ ($2*(n.south)$) -- ++ ($0.2*(n.west)$);
  }}  

\newcommand\id[1]{\mathds{1}_{#1}}

\usepackage{textgreek}
\newcommand{\ppi}{\text{\textpi}}

\DeclareMathAlphabet{\mathpzc}{OT1}{pzc}{m}{it}

\newcommand\ii{\mathrm{i}}
\newcommand\dd{\mathrm{d}}
\newcommand\ee{\mathrm{e}}
\newcommand\norm[1]{\left\| #1 \right\|}
\newcommand\mat[1]{\big[ #1 \big]}
\newcommand\minus{\! \scalebox{0.7}[1.0]{\( - \)}}
 
\providecommand{\tprod}[2]{\left(#1 \otimes #2\right)} 
\newcommand\mycomment[1]{}

\title{Exact Functional Integration of Radial and Complex Slave-boson Fields~:  
Thermodynamics and Dynamics of the Two-site Extended Hubbard Model} 
\author[V. H. Dao]{Vu Hung Dao\inst{1}}
\author[R. Fr\'esard]{Raymond Fr\'esard\inst{1,}\footnote{Corresponding
    author\quad 
			E-mail:~\textsf{Raymond.Fresard@ensicaen.fr},
			Phone: +33\,231\,45\,26\,09,
			Fax: +33\,231\,95\,16\,00}}
\address[1]{Normandie Universit\'e, ENSICAEN, UNICAEN, CNRS, CRISMAT, 14000
  Caen, FRANCE} 
\shortauthors{V. H. Dao et al.}

\begin{abstract}
The functional integral formulation of the Hubbard model when treated in its 
Kotliar-Ruckenstein representation in the radial gauge involves fermionic, 
as well as complex and radial slave boson fields. In order to improve on the 
understanding of the interplay of the three types of fields, and on the nature 
of the latter, we perform a comprehensive investigation of an exactly solvable 
two-site cluster, as it entails all pitfalls embodied in this approach. It is 
first shown that the exact partition function is recovered, even when 
incorporating in the calculation the square root factors that are at the heart 
of the representation, when suitably regularized. 
We show that using radial slave boson fields allows to overcome all hurdles 
following from the normal ordering procedure. We then demonstrate that this 
applies to the Green's function as well, and to the correlation functions of 
physical interest, thereby answering the criticisms raised by Sch{\"o}nhammer 
[K. Sch{\"o}nhammer, Phys. Rev. B \textbf{1990} \textit{42}, 2591]. In addition, the 
investigation generalizes the calculations to the Hubbard model extended by 
a non-local Coulomb interaction.
\end{abstract}
\shortabstract
\begin{document}
\maketitle

\section{Introduction} 
Strongly correlated electrons on a lattice, as encountered in transition-metal
oxides, occupy a prominent place among the most interesting and challenging 
topics of contemporary theoretical physics. This interest is largely fueled by 
the functionality-oriented properties these systems harbor, such as high-T$_c$ 
superconductivity (see, e.g.,~\cite{Bed86,Raveau91,Malozemoff05}), colossal 
magnetoresistance (see, e.g.,~\cite{Helmolt93,Tomioka95,Raveau95,Maignan95}), 
transparent conducting oxides (see, e.g.,~\cite{Kawa97}), high capacitance 
heterostructures~\cite{Li11} or large thermopower (see, 
e.g.,~\cite{Ter97,Mas00,Mat01,Mic07,Ohta07,Wang13,Gui11}), to quote a few. 
Furthermore, strongly correlated transition-metal
oxides comprise rare-earth free permanent 
magnets~\cite{Pullar2012}, materials for batteries~\cite{Mizushima1980,
Nagaura1990,Arico2005,Kang2006}, and multiferroics~\cite{Poienar2009,Maignan2018,Wang2003}.
Moreover, a broad diversity of further interesting properties are
exhibited by these systems. They involve correlation-driven metal-to-insulator
transition at the first place, for instance in vanadium
sesquioxide~\cite{McW73,Hel01,Lim03} or in hole-doped titanates~\cite{Tokura1993}, 
as well as a rich palette of ordered phases of, e.g., magnetic, charge, stripes, 
nematic and, above all, superconducting nature. 

The Hubbard model is the simplest model Hamiltonian entailing a genuine
competition between the kinetic energy of electrons hopping on a lattice and
the Coulomb interaction energy simplified to its sole local
component. Initially introduced, \textit{inter alia}, to describe metallic
magnetism~\cite{Hub63,Kan63,Gutz63}, it regained an immense popularity after 
Anderson's proposal that the Hubbard model on the square lattice entails the
key ingredients to high-T$_c$ superconductivity~\cite{And87}, which 
$d$-wave symmetry may hardly be grasped in a simple one-electron picture. 
As of today, considering longer-ranged interaction is receiving increasing 
attention~\cite{Deeg1993,Aichhorn2004,Davoudi2006,Kagan2011,Ayral2013,vanLoon2014,
Lhoutellier2015,Steffen2017,Terletska2017,Kapcia2017,Schuler2018,Paki2019,
Terletska2021,Roig2022,Philoxene2022,Linner2023,Riegler2023,AucarBoidi2023}
(see also \cite{Kundu2023} and references therein for a more complete list).

Strongly correlated electron systems pose challenges that have been tackled 
using slave-boson approaches in a number of fashions. They mostly back on 
Barnes'~\cite{Barnes76,Barnes77} and Kotliar and Ruckenstein's (KR)~\cite{KR} 
representations, as well as on their multiband and rotation invariant
generalizations~\cite{FK,LiWH,FW92,Riegler20,Kotliar-Georges,Piefke}. They 
all exhibit their own gauge symmetry group.

As regards Barnes' representation to the infinite-$U$ single-impurity Anderson
model, it involves one doublet of fermions, one slave boson, and one
time-independent constraint binding all three fields. It entails a $U(1)$
gauge symmetry, that may be used to gauge away the phase of the slave boson. 
This, however, requires to introduce a time-dependent constraint 
field~\cite{RN83a,RN83b,NR87}. The argument was originally put forward in the 
continuum limit, but later, a path integral representation on discrete-time 
steps for the remaining radial slave boson together with the time-dependent 
constraint has been proposed by one of us~\cite{RFTK01}. It could be used to
solve the Ising chain~\cite{RFTK01}.

A deeper understanding of radial slave bosons was gained through an 
analysis of a toy model. It yielded the exact expectation value of
the radial slave boson to be generically finite~\cite{RFHOTK07}, as well as a
way to exactly handle functional integrals involving constrained fermions and
radial slave bosons. A further feature of radial slave bosons is related to
the saddle-point approximation. When the latter is performed to a complex slave
boson field, it is intimately tied to a Bose condensation, that is generally
viewed as spurious. On the contrary, radial slave bosons are phase-stiff and
do not Bose condense, as they are deprived of a phase degree of
freedom. Accordingly, their saddle-point amplitude approximates their --
generically finite -- exact expectation value. 

The KR representation~\cite{KR}, and related slave-boson
representations~\cite{LiWH,FW92,Riegler20}, have first been set up to the
investigation of the Hubbard model, and have then been extended to account 
for long-ranged density-density and spin-spin interactions~\cite{Lhoutellier2015}.
Meanwhile, a whole range of valuable results have been obtained, primarily on 
the square lattice. 
More specifically, they have been applied to the Mott metal-to-insulator 
transition~\cite{Vollhardt1987,Dao17,Mezio2017}, then to the description of 
anti-ferromagnetic~\cite{Lil90}, spiral~\cite{Fre91,Igo13,Fre92,Doll2}, 
striped~\cite{SeiSi,Fle01,Sei02,Rac06a} phases, and even to the
competition between the latter two has been addressed~\cite{RaEPL}. In
addition, spin-and-charge ordered phases of a half-filled extended Hubbard
model, which may be suitable for resistive switching, were recently put 
forward~\cite{Philoxene2022}. Besides, it has been obtained that the spiral order 
continuously evolves to the ferromagnetic order in the large $U$ regime 
($U \gtrsim 60t$)~\cite{Doll2} so that it is unlikely to be realized 
experimentally. Consistently, in the two-band model, ferromagnetism was
found as a possible groundstate in the doped Mott insulating 
regime~\cite{Fre02}, only. Yet, the ferromagnetic instability line may be 
brought down to the intermediate coupling regime in various ways. They involve 
adding a nearest-neighbor ferromagnetic exchange coupling~\cite{Lhoutellier2015}
or a sufficiently large next-nearest-neighbor hopping amplitude~\cite{FW98}, to
quote a few. Going to the fcc lattice~\cite{Igo15} results in the same effect.

The KR representation implies six auxiliary particles~\cite{KR}. Two of them
form a doublet of fermions, the remaining ones are bosons, and they are
``enslaved'' by three time-independent constraints. Ending a long-standing
debate~\cite{Ras88,Lil90,Lav90,Li94}, a consensus that the gauge symmetry
group reads $U(1)\times U(1) \times U(1)$ has been
reached~\cite{Jol91,FW92,Kot92}. Accordingly, the phase of three of the four
slave bosons can be gauged away, with the counterpart that all three
constraints become time-dependent constraint fields. Therefore one bosonic
field remains as a complex one. It may be chosen to be the $d$ field, which 
accounts for doubly occupied sites, or the $e$ field, which accounts for empty 
sites. These two fields are at the first place associated with charge
fluctuations, and one might wonder whether the charge fluctuation spectrum
depends on this asymmetry, but it could be shown that this is not the case, at
least when it is computed to one-loop order around the paramagnetic 
saddle-point~\cite{Dao17,Dao18}. In this context this one-loop calculation 
was believed to coincide with the time-dependent Gutzwiller approximation 
(TDGA) until discrepancies between the two were put forward~\cite{Dao17}. 
This motivated an extension of the TDGA to achieve agreement with the one-loop
calculation~\cite{Noatschk2020}.

A whole series of functional integral representations has been formulated 
for various correlated systems. They quite systematically make use of coherent 
states~\cite{Perelomov1986}. This also applies to the current auxiliary fermionic
and complex bosonic fields, but not to the radial slave bosons for which the 
functional integral representation under study is established from the outset.
By now, the received attention has been limited to the atomic limit~\cite{Dao20}, 
and the purpose of this work is to establish the handling of the delocalization 
of the electrons due to hopping. 
To that aim, we compute in this context the partition function
and dynamical quantities of the extended Hubbard model on the two-site cluster, 
and we show that the exact results  are recovered for all densities. Eventually, 
we furthermore demonstrate that the so-called Kotliar roots entering the kinetic 
energy may equally well be included in the calculation, provided a new 
regularization scheme is enforced, with the same result. We thereby generalize 
earlier works where all relevant slave-boson fields entered as radial 
fields~\cite{RFTK01,RFHOTK07,RFTK12}. 

The paper is organized as follows. In Section~\ref{sec:model} we present the 
model and its representation in terms of KR slave bosons in the radial gauge. 
The functional integration of the latter is further detailed in 
Section~\ref{sec:func_int}. In order to validate the procedure and to illustrate 
the key features of the computation, the paper presents the evaluation of the 
partition function, of the physical electron Green's function, and of several 
thermal averages of quantities represented by radial slave bosons.
The fermionic fields are integrated out in Section~\ref{sec:int_f}, where $N$-time 
products are introduced. The integration over the complex bosonic field 
is performed in Section~\ref{sec:int_d}, and the remaining ones in 
Section~\ref{sec:int_P}. 
The agreement between the so-computed thermal averages and the expressions 
derived from the physical electron Hamiltonian is established in 
Section~\ref{sec:recovering}. 
Section~\ref{sec:conclusion} summarizes our work. 
The calculation of the pseudofermion Green's function is extensively presented 
in Appendix~\ref{app:det_min}, while the irregular contributions to the traces 
involved in the computation of thermal averages are explicitly handled in 
Appendix~\ref{app:limit_irr}.

\section{The model and its Kotliar and Ruckenstein representation}\label{sec:model}
We investigate the extended Hubbard model in an exactly soluble case, which is 
the two-site cluster. It reads
\begin{align}\label{eq:hamilton}
 {\cal H}  =  & \sum_{\sigma}  \sum_{i=1}^{2} \left( \varepsilon 
 c_{\sigma,i}^{\dagger}  c_{\sigma,i}^{\phantom{\dagger}} 
 - t c_{\sigma,i}^{\dagger}c_{\sigma,i-1}^{\phantom{\dagger}} \right) 
  + U \sum_{i=1}^{2}  c_{\uparrow,i}^{\dagger} c_{\uparrow,i}^{\phantom{\dagger}} 
 c_{\downarrow,i}^{\dagger} c_{\downarrow,i}^{\phantom{\dagger}} 
 \nonumber \\
 & + V  \prod_{i=1}^2 \left( c_{\uparrow,i}^{\dagger} c_{\uparrow,i}^{\phantom{\dagger}}
 +  c_{\downarrow,i}^{\dagger} c_{\downarrow,i}^{\phantom{\dagger}} \right),
\end{align}
and is made of the single-electron Hamiltonians for each of the spin projections
($\sigma=\uparrow,\downarrow$), supplemented by the Coulomb interaction terms, 
with amplitude $U$ on each site $i$, and $V$ on the bond.
The factor $\varepsilon$ is the difference in energy between the orbital level and 
the chemical potential. In the hopping terms with amplitude $t$, we use the periodic 
boundary condition in order to shorten the notation.  

The Kotliar and Ruckenstein (KR) representation~\cite{KR} of this model involves one 
doublet of fermionic fields $\{ f_{\uparrow}, f_{\downarrow} \}$ and four bosonic
fields $\{e, p_{\uparrow}, p_{\downarrow},d\}$. 
The latter are tied to an empty site, single occupancy of the site 
with spin projection up or down, and double occupancy, respectively.
This implies redundant degrees of freedom which have to be discarded. 
This is achieved by imposing the three constraints
\begin{subequations} \label{eq:q2_KR}
\begin{eqnarray}
e^\dagger e^{\phantom{\dagger}} + \sum_\sigma p^{\dagger}_{\sigma}p^{\phantom{\dagger}}_{\sigma}
+ d^\dagger d^{\phantom{\dagger}} &=&1 \\
p^{\dagger}_{\sigma}p^{\phantom{\dagger}}_{\sigma} 
+ d^\dagger d^{\phantom{\dagger}} &=& f^{\dagger}_{\sigma} f^{\phantom{\dagger}}_{\sigma}
\;\;\;\;\;\;\;\;\;\;(\sigma =\uparrow,\downarrow)
\end{eqnarray}
\end{subequations}
to be satisfied on each site, thereby enslaving the bosonic fields.

The Hamiltonian in the so enlarged Fock space is 
\begin{align}\label{H_KR}
{\cal H}_{\rm KR} = &   \sum_{\sigma} \sum_{i=1}^{2}  \left( \varepsilon 
 f_{\sigma,i}^{\dagger}  f_{\sigma,i}^{\phantom{\dagger}} 
 - t f_{\sigma,i}^{\dagger} z_{\sigma,i}^{\dagger}z_{\sigma,i-1}^{\phantom{\dagger}} 
 f_{\sigma,i-1}^{\phantom{\dagger}} \right) 
  \nonumber \\
 & + U \sum_{i=1}^{2}  d_i^{\dagger} d_i^{\phantom{\dagger}} 
  + V \prod_{i=1}^2 \left( 2 - 2 e_{i}^{\dagger} e_{i}^{\phantom{\dagger}} 
  - \sum_{\sigma} p_{\sigma,i}^{\dagger} p_{\sigma,i}^{\phantom{\dagger}} \right)
\end{align}
where
\begin{equation}
z_{\sigma,i} =   e_i^{\dagger} L_{\sigma,i}^{\phantom{\dagger}} R_{\sigma,i}^{\phantom{\dagger}}
p_{\sigma,i}^{\phantom{\dagger}} + p_{-\sigma,i}^{\dagger}  L_{\sigma,i}^{\phantom{\dagger}} 
R_{\sigma,i}^{\phantom{\dagger}} d_i^{\phantom{\dagger}}
\end{equation}
is the occupancy change operator. The latter is renormalized by inverse square root factors
\begin{subequations}\label{eq:K_roots}
\begin{align}
 L_{\sigma,i}^{\phantom{\dagger}} & = \left(1 - p_{\sigma,i}^{\dagger} p_{\sigma,i}^{\phantom{\dagger}} 
- d_i^{\dagger} d_i^{\phantom{\dagger}}\right)^{-\frac{1}{2}} \\
 R_{\sigma,i}^{\phantom{\dagger}} & = \left(1 - e_i^{\dagger} e_i^{\phantom{\dagger}} 
- p_{-\sigma,i}^{\dagger} p_{-\sigma,i}^{\phantom{\dagger}} \right)^{-\frac{1}{2}} 
\end{align}
\end{subequations}
These so-called 'Kotliar roots' modify the saddle-point approximation so that it 
yields the Gutzwiller approximation result~\cite{KR}. Furthermore, they act exactly 
as the identity operator within the physical subspace, as we show below. 
Yet, as emphasized by Sch\"onhammer~\cite{Schonhammer1990}, their exact computation 
within the functional integral formalism poses a formidable challenge, as there was 
not any rigorous procedure to substitute these functions of boson operators 
in the Lagrangian by field expressions, after the Hamiltonian is written in the 
normal order form~\cite{negele}. 
This step is usually neglected since, by doing so, the Gutzwiller approximation is 
recovered as a saddle-point, which is variationally controled in the large-dimensionality 
limit~\cite{Metzner1988,Metzner1989a,Metzner1989b}.
The purpose of the present paper is then to remedy this loophole, and to show that 
the radial gauge KR representation, with a proper regularization scheme, 
allows to overcome these mathematical hurdles when exactly evaluating the Kotliar 
roots within the functional integrals.

 In this formalism the physical constraints~(\ref{eq:q2_KR}) are 
 enforced by the three Lagrange multipliers $\lambda_i$, resp. $\lambda_{\sigma,i}$, 
 associated to each site. 
 At this stage it should be noted that the functional integral over the fermionic and 
 bosonic fields cannot be performed right away. Indeed, in order to ensure convergence, 
 $\lambda_i$ has to be continued into the complex plane as 
\begin{equation}
\tilde{\lambda}_i = \lambda_i - \ii \lambda_0  
\end{equation}
with $\lambda_0 > 0$ (or $\lambda_0 + U >0$ if $U<0$) so that the integration 
contour is shifted into the lower half-plane \cite{Bickers86,Bickers87,RFTK01}.

The expression of the Lagrangian in the Cartesian representation of the 
slave boson fields reads
\begin{equation}\label{eqLtot}
{\mathcal{L}}^{(c)}(\tau) = {\mathcal{L}}^{(c)}_f(\tau) + {\mathcal{L}}^{(c)}_b(\tau). 
\end{equation}
It entails the dynamics of the auxiliary fermionic and bosonic fields, 
together with the constraints~(\ref{eq:q2_KR}) specific to the KR setup. 
The fermionic contribution is quadratic in the fermion fields with
\begin{align}\label{Lcont_f}
{\mathcal{L}}^{(c)}_f (\tau) =& \sum_{\sigma} \sum_{i=1}^2 \left(
f^{\ast}_{\sigma,i}(\tau) \left(\partial_{\tau} + \epsilon  +
\ii \lambda_{\sigma,i}\right)f^{\phantom{\dagger}}_{\sigma,i}(\tau) \right. \nonumber \\
 & -t  \left. f^{\ast}_{\sigma,i}(\tau) z_{\sigma,i}^{*}(\tau) 
 z_{\sigma,i-1}^{\phantom{*}}(\tau) f^{\phantom{\dagger}}_{\sigma,i-1}(\tau)  \right).
\end{align}
The remaining part
\begin{eqnarray}\label{Lcont_b}
{\mathcal{L}}^{(c)}_b(\tau) &=& \sum_{i=1}^2 \bigg( - \ii\tilde{\lambda}_i  + e^{*}_i(\tau) 
\left(\partial_{\tau} + \ii \tilde{\lambda}_i\right) e_i(\tau)  
\nonumber\\
&\quad&  \quad   + d^{*}_i(\tau) \left(\partial_{\tau} + U + \ii\tilde{\lambda}_i 
- \ii \lambda_{\uparrow,i} - \ii \lambda_{\downarrow,i}\right) d_i(\tau)  
\nonumber \\
&\quad & \quad + \sum_{\sigma} p^{*}_{\sigma,i}(\tau) \left(\partial_{\tau} 
+ \ii\tilde{\lambda}_i - \ii \lambda_{\sigma,i} \right) p^{\phantom{*}}_{\sigma,i}(\tau) 
\bigg) 
\nonumber \\
& \quad &  + V \prod_{i=1}^2 \left( 2 
- 2 e_{i}^{*}(\tau) e_{i}^{\phantom{*}}(\tau) 
- \sum_{\sigma} p_{\sigma,i}^{*}(\tau) p_{\sigma,i}^{\phantom{*}}(\tau) 
\right)
\end{eqnarray}
 is only composed with boson fields, up to the constant terms. And we get the partition 
 function in the Cartesian gauge as
 \begin{align}\label{eq:Z_cart}
  {\cal Z} &=   \int_{-\ppi/\beta}^{\ppi/\beta} \prod_{i=1}^2 \left(
  \frac{\beta \, \dd\!\lambda_i}{2\ppi}  \prod_{\sigma} 
  \frac{\beta\, \dd\!\lambda_{\sigma,i}}{2\ppi} \right)\; \int\! \prod_{\sigma} 
  {\rm D}\big[f_{\sigma}^{\phantom{\dagger}}, f_{\sigma}^{\ast} \big]  
  \nonumber \\
  &\times \int \!\! {\rm D}\big[e, e^{*} \big] 
  \,{\rm D}\big[d, d^{*} \big] 
  \prod_{\sigma}  {\rm D}\big[p_{\sigma}^{\phantom{*}}, p_{\sigma}^{*} \big] 
  \;\ee^{-\int_0^{\beta} d\tau {\cal L}^{(c)} (\tau) }  .
 \end{align}
 
In this representation, the physical electron creation and annihilation
operators are mapped onto auxiliary operators as
\begin{equation}\label{eq:KR}
          c_{\sigma,i}^{\phantom{\dagger}} \mapsto
          z_{\sigma,i}^{\phantom{\dagger}} f_{\sigma,i}^{\phantom{\dagger}} 
          \quad\quad {\rm and} \quad\quad 
          c_{\sigma,i}^{\dagger} \mapsto
          f_{\sigma,i}^{\dagger} z_{\sigma,i}^{\dagger} .
\end{equation}
They properly anticommute provided the constraints~(\ref{eq:q2_KR}) are satisfied.
This representation of the physical electron operators is invariant under the 
 gauge transformations
\begin{equation}
  \left\{ \begin{array}{l}
          f_{\sigma,i} \mapsto \ee^{-\ii \chi_{\sigma,i}} 
          f_{\sigma,i}
          \\
          e_i \mapsto \ee^{\ii \theta_i} e_i
          \\
          p_{\sigma,i}\mapsto \ee^{\ii \left(\chi_{\sigma,i} + \theta_i \right)}
          p_{\sigma,i}
          \\
          d_i \mapsto \ee^{\ii \left(\chi_{\uparrow,i} + 
          \chi_{\downarrow,i}  + \theta_i\right)} d_i.
         \end{array} 
   \right.      
\end{equation}
The local gauge symmetry group is therefore $U(1) \times U(1) \times U(1)$ on each site.
The Lagrangian, Eq.~(\ref{eqLtot}), also possesses this symmetry. 
Expressing the bosonic fields in amplitude and phase variables as
\begin{subequations} 
\begin{equation}
e_i(\tau)  =  \sqrt{R_{e,i}(\tau)}\, \ee^{\ii \theta_i(\tau)} 
\end{equation}
\begin{equation}
p^{\phantom{\dagger}}_{\sigma,i} (\tau)  =  \sqrt{R_{\sigma,i}(\tau)}\,
\ee^{\ii (\chi_{\sigma,i}(\tau) + \theta_i(\tau))} 
\end{equation}
\end{subequations}
allows to gauge away the phases of three of the four slave boson fields 
provided one introduces the three time-dependent Lagrange multipliers
\begin{subequations}
\begin{eqnarray}
\alpha_i (\tau) & = &\lambda_i + \partial_{\tau}
\theta_i(\tau) \\
\beta_{\sigma,i} (\tau) & = & \lambda_{\sigma,i}
  - \partial_{\tau} \chi_{\sigma,i}(\tau) .
\end{eqnarray}
\end{subequations}
Here the radial slave boson fields are implemented in the continuum limit
following, e.~g., Ref.~\cite{RN83a,RN83b,NR87}. Similar gauge symmetry
groups have been identified in the spin rotation invariant 
representation~\cite{FW92} and in the case of the two-band model~\cite{RFKSTK22}.

\section{Functional integrals in the radial gauge}\label{sec:func_int}

For the exact evaluation of the functional integrals, the representation in 
the radial gauge has to be set up on a discretized time mesh from the beginning. 
Moreover, the constraints now have to be satisfied at every time step.
Extending the procedure introduced in Ref.~\cite{RFTK01} for Barnes' slave 
boson to the KR representation one may compute the thermal average $\big\langle 
\mathscr{Q} \big\rangle$ of a quantity $ \mathscr{Q} $ as
\begin{strip}
  \begin{align}\label{eq:th_average}
  {\cal Z} \big\langle \mathscr{Q} \big\rangle =  \lim_{\nu\to 0} \lim_{N\to \infty} \lim_{\eta\to 0^{+}} 
  & \left\{ \prod_{n=1}^N \prod_{i=1}^2 \left(  \int_{-\eta}^{\infty}  
   \dd\! R_{e,i,n}\;  \dd\! R_{\uparrow,i,n}\; \dd\! R_{\downarrow,i,n} 
   \int_{-\infty}^{\infty}   \frac{\delta \dd \alpha_{i,n}}{2 \ppi}
 \frac{\delta \dd \beta_{\uparrow, i,n}}{2 \ppi} 
 \frac{\delta \dd \beta_{\downarrow, i,n}}{2 \ppi} 
 \int \!\!  \frac{\dd d_{i,n}^{\phantom{*}}\; 
 \dd d_{i,n}^{*}}{2\ppi\ii} \int \prod_{\sigma} \dd\!f_{\sigma,i,n}^{\phantom{\ast}} 
 \; \dd\! f_{\sigma,i,n}^{\ast}  \right) \right.
 \nonumber \\
 & \left.   \phantom{\frac{d_n^*}{\ppi}}\!\!\!\!\!\! \times
  Q \; \ee^{-S} \; \right\}
 \end{align}
where $Q$ is its discrete-time representation, and the action reads
\begin{align}\label{eq:S_radial}
 S = &\sum_{n=1}^N \left\{  \sum_{i=1}^2 \Bigg[ \sum_{\sigma} 
f_{\sigma,i,n}^{\ast}\left(f_{\sigma,i,n}^{\phantom{\ast}} - 
\ee^{-\delta( \epsilon  + \ii \beta_{\sigma,i,n})} 
f_{\sigma,i,n-1}^{\phantom{\ast}} \right. \right. 
 + \delta t  \left.z_{\sigma,i,n}^{\star}  z_{\sigma,i-1,n-1}^{\phantom{*}} 
 f^{\phantom{\dagger}}_{\sigma,i-1,n-1}  \right)
 - \ii\delta \tilde{\alpha}_{i,n}
 + \ii\delta  \sum_{\sigma} \left(\tilde{\alpha}_{i,n} - \beta_{\sigma,i,n} \right) 
  R_{\sigma,i,n}  
 \nonumber \\
& \quad\quad\quad + \ii\delta \tilde{\alpha}_{i,n} R_{e,i,n}  + d^{*}_{i,n} \left( d_{i,n}^{\phantom{*}}- \ee^{-\delta (U + \ii
\tilde{\alpha}_{i,n} - \ii\beta_{\uparrow,i,n} - \ii\beta_{\downarrow,i,n})} 
d_{i,n-1}  \right) \Bigg]
+ \left. \delta V \prod_{i=1}^2 \left( 2 - 2 R_{e,i,n} 
- \sum_{\sigma} R_{\sigma,i,n} \right) \right\}.
\end{align}
\end{strip}
Here the integer $N$ is the number of imaginary-time slices with duration
\begin{equation}
\delta = \frac{\beta}{N}  
\end{equation} 
while $\nu$ and $\eta$ are regulators, the purposes of which will 
be explained below.   
The integral expression of the partition function $\cal Z$ is obtained with
$\mathscr{Q}= Q =1$. Note that we have introduced the shorthand notation
\begin{equation}
\tilde{\alpha} \equiv \alpha - \ii \lambda_0 . 
\end{equation}
As previously discussed, the functional integral has to be evaluated with $\alpha$ 
replaced by ($\alpha - \ii \lambda_0$) with $\lambda_0 > 0$, in order to ensure 
convergence.  

The functional integral Eq.~(\ref{eq:th_average}) may be equivalently recast
in the more suggestive formulation
 \begin{align}\label{eq:th_average2}
  {\cal Z} \big\langle \mathscr{Q} \big\rangle &= 
  \lim_{\nu\to 0} \lim_{N\to \infty}
   {\cal P} \left( \big\langle Q \big\rangle_{f_{\uparrow}, 
   f_{\downarrow}, d} \right).
 \end{align}
 Here the  operator ${\cal P}$, which will be detailed below, 'projects onto the 
 physical subspace' the  correlation
 $\big\langle Q \big\rangle_{f_{\uparrow}, f_{\downarrow}, d}$ 
 that is obtained as the thermal average of $Q$ over the different 
 configurations of the pseudofermion and $d$-boson fields within the enlarged Fock 
 space~: It discards all the contributions that do not comply with the 
 constraints~(\ref{eq:q2_KR}), while  properly weighting the remaining ones.
 
The partition function may thus be computed as
\begin{equation}\label{eqh2}
\mathcal{Z}  = \lim_{\nu\to 0} \lim_{N\to \infty}
{\mathcal P} \left(  Z_{df} \right)
\end{equation}
where the joint partition function $Z_{df}$ of the $d$ boson and the auxiliary fermions
is evaluated as the functional integral
\begin{align}
 Z_{df}=   \big\langle  1 \big\rangle_{f_{\uparrow},f_{\downarrow},d}
\end{align}
with
\begin{subequations}
\begin{align}
 \big\langle \dots \big\rangle_{f_{\sigma}} &\equiv
 \int \!\!  \prod_{n=1}^N \prod_{i=1}^2 \dd\!f_{\sigma,i,n}^{\phantom{\dagger}} 
 \; \dd\! f_{\sigma,i,n}^{\ast}
 \; \dots  \;  \ee^{-S_{\sigma}},
 \\
 \big\langle \dots \big\rangle_{\! d} &\equiv
 \int \!\!  \prod_{n=1}^N \prod_{i=1}^2 \frac{\dd d_{i,n}^{\phantom{*}}\; 
 \dd d_{i,n}^{*}}{2\ppi\ii} 
 \dots  \;  \ee^{-S_{d}} ,
\end{align}
\end{subequations}
where $S_{\sigma}$ and $S_{d}$ are the terms in the action~(\ref{eq:S_radial}) that are 
quadratic in the variables $f_{\sigma,i,n}$ and $d_{i,n}$, respectively. 

The fermionic action $S_f =  S_{\uparrow} + S_{\downarrow}$ is the sum of 
the quadratic forms represented by the $2N \times 2 N$ matrix
\begin{align}\label{eq:s_sigma}
\big[ S_{\sigma} \big]
=  \begin{bmatrix}
     \id{2} &  &  & -\big[M_{\sigma,1}\big] \\
     \big[M_{\sigma,2}\big] & \id{2}  &  & \\
      & \ddots  &  \ddots &   \\
      & & \big[M_{\sigma,N}\big] & \id{2} 
    \end{bmatrix}
\end{align}
within the basis $\{ f_{\sigma,1,1}, f_{\sigma,2,1}, \ldots,  f_{\sigma,1,N}, f_{\sigma,2,N}\}$
 for each spin projection. Here $\id{2}$ is the $2\times2$ identity matrix, and 
 the block 
\begin{equation}
 \big[M_{\sigma,n}\big] = \begin{bmatrix}
                   - L_{\sigma,1,n} & & T_{\sigma,1,n} \\
                    T_{\sigma,2,n} & &- L_{\sigma,2,n}
                  \end{bmatrix}
\end{equation}
involves
\begin{subequations}
\begin{align} 
 \label{mat_elem_L} L_{\sigma,i,n} & =  \ee^{-\delta(\varepsilon + \ii \beta_{\sigma,i,n})}  \\
 \label{mat_elem_T} T_{\sigma,i,n} & = \delta t z_{\sigma,i,n}^{\star} 
 z_{\sigma,i-1,n-1}^{\phantom{\star}}.
\end{align}
\end{subequations}
The representations in the radial gauge of the 
operators $z_{\sigma,i}$ and $z_{\sigma,i}^{\dagger}$ are, respectively,
\begin{align} \label{eq:z_sin}
  z_{\sigma,i,n} &=
\frac{ \sqrt{R_{e,i,n+1} R_{\sigma,i,n}} + \sqrt{R_{-\sigma,i,n+1}} d_{i,n} }
{\sqrt{ R_{e,i,n+1} + R_{-\sigma,i,n+1} - \ii \nu } 
\sqrt{ 1 - R_{e,i,n} - R_{-\sigma,i,n} + \ii \nu } }\nonumber \\
  {z}_{\sigma,i,n}^{\star} &=
\frac{ \sqrt{R_{\sigma,i,n+1} R_{e,i,n} } + d_{i,n}^{*} \sqrt{R_{-\sigma,i,n}}  }
{\sqrt{ 1 - R_{e,i,n+1} - R_{-\sigma,i,n+1} - \ii \nu } 
\sqrt{R_{e,i,n} + R_{-\sigma,i,n} + \ii \nu } }
\end{align}
where the radial variable $R_{e,i, n}$ ($R_{\sigma,i, n}$) corresponds to the squared
amplitude of the complex $e_{i,n}$ ($p_{\sigma,i, n}$) bosonic field. First, note
that ${z}_{\sigma,i,n}^{\star}$ is not the complex conjugate of $z_{\sigma,i,n}$
as the time steps of radial variables are not the same. Second, the above expressions 
of the $z$ factors differ from the usual ones. Indeed, we
made use of the physical constraints~(\ref{eq:q2_KR}) to replace the number of $d$ bosons
in the denominators by its counterpart in terms of radial slave-boson variables, 
which eases the evaluation of the functional integrals. 
The regulator $\nu$ is taken to zero after performing the continuous-time limit 
$N\rightarrow \infty$. It is introduced in the discrete-time representation in order to 
take care of the vanishingly small contribution of spurious processes which appear when 
computing the partition function, as it ensures that the $z$ factors are not singular in 
the physical subspace.

The part of the action $S_d = S_{d_1} + S_{d_2} $ that is quadratic in the $d$ boson 
field, is the sum of the contributions from each site
\begin{align}\label{eq:sd}
S_{d_i}
&=  \sum_{m=1}^N \sum_{n=1}^N d_{i,m}^{*} \big[S_{d_i}\big]_{m,n}^{\phantom{*}} d_{i,n}^{\phantom{*}}
\nonumber
\\
&= \sum_{n=1}^{N} 
d_{i,n}^{*} \Big( d_{i,n}^{\phantom{*}} - d_{i,n-1}^{\phantom{*}} \ee^{\delta(- U - \ii \tilde{\alpha}_{i,n} 
+ \ii \beta_{\uparrow, i,n } + \ii \beta_{\downarrow,i, n }) } \Big)
\end{align}
where $d_{i,0} \equiv d_{i,N}$ in the second line to satisfy periodic boundary 
conditions in the imaginary time. When expanding the exponential in Eq.~(\ref{eq:sd}) 
to lowest order in $\delta$, the familiar form following from the Trotter-Suzuki 
decomposition is recovered. Yet, the latter may only be applied for bounded values of 
the Lagrange multipliers while Eq.~(\ref{eq:sd}) is well behaved when integrating the 
multipliers along the real axis. 

The remaining terms of the action are gathered in the integral operator
\begin{equation}\label{eq:projp}
{\mathcal P} \equiv \prod_{n=1}^{N}  {\mathcal P}_{n}
\end{equation}
where
\begin{align}\label{eq:projp2}
{\mathcal P}_{n} \equiv & \int_{\! R_n} 
 \!\!\ee^{- \delta V_n} 
\int_{\alpha \beta_n}
\!\!\ee^{\ii\delta \sum\limits_{i=1}^2 \left(  \tilde{\alpha}_{i,n}
(1 - R_{e,i,n} - \sum\limits_{\sigma} R_{\sigma, i,n}) 
+ \sum\limits_{\sigma} \beta_{\sigma,i, n} R_{\sigma,i, n}\right)} 
\end{align}
is defined with the non-local Coulomb potential
\begin{equation}
 V_n =  V \prod\limits_{i=1}^2 \left( 2 - 2 R_{e,i,n}  
 - R_{\uparrow,i,n} - R_{\downarrow,i,n} \right) 
\end{equation}
and the shorthand notations
\begin{subequations}
\begin{align}
 \int_{ R_n} \;  & \equiv \;  \lim_{\eta\to 0^{+}}
\int_{-\eta}^{\infty} \prod_{i=1}^2 
\left\{ \dd\! R_{e,i,n}\;  \dd\! R_{\uparrow,i,n}\; \dd\! R_{\downarrow,i,n} \right\}, 
 \\
 \int_{\alpha \beta_n} 
& \equiv \; \int_{-\infty}^{\infty} \prod_{i=1}^2 \left\{ \frac{\delta \dd \alpha_{i,n}}{2 \ppi}
\frac{\delta \dd \beta_{\uparrow, i,n}}{2 \ppi} 
\frac{\delta \dd \beta_{\downarrow, i,n}}{2 \ppi} \right\}.
\end{align}
\end{subequations}
Note that it takes an infinitesimal regulator $-\eta$ for the integration bounds to 
have well defined delta functions enforcing the constraints.

The integration over the fields will be carried out in the following order. First, 
integrating the fields $f_{\sigma}$ and $d$ yields the joint partition function 
$Z_{df}$ as a sum over all the discrete-time evolutions that are governed by 
the KR Hamiltonian~(\ref{H_KR}). Note that at this stage, the dynamics is not 
restricted to the subspace of physical states. Then, we perform the integration over  
the Lagrange multipliers and the radial fields through the operator ${\cal P}$, 
Eq.~(\ref{eq:projp}). 
This removes the contributions of the unphysical trajectories in the evaluation 
of the partition function ${\cal Z}$, and it multiplies the physical ones by their 
respective weight factor associated with the non-local Coulomb interaction energy. 
The latter integrations are straightforward for static settings where no hopping can 
occur. The computation is more involved in the presence of hopping processes because 
the boson number distribution is modified on all relevant pairs of sites. 
The calculations prove that ${\cal P}$ does filter out all irrelevant 
evolutions as intended, so that the remaining contributions yield the expected expression
of the partition function in the continuous-time limit.

Other thermal averages may be integrated in the same fashion. The most 
similar computation certainly is evaluating a pure correlation of radial 
boson fields since for any function ${\cal F}$ of only radial variables, 
\begin{equation}
\big\langle {\cal F}(R_a, R_b, \dots) 
\big\rangle_{f_{\uparrow},f_{\downarrow},d}
= {\cal F}(R_a, R_b, \dots) Z_{df}. 
\end{equation}
A whole range of static thermal averages and correlation functions may be 
expressed in terms of them. Let us focus here on the $R_e$ field as an example, with 
the evaluation of the hole density $\left\langle R_{e,1} \right\rangle$ on site 1 
given by 
\begin{equation}
 {\cal Z} \left\langle R_{e,1} \right\rangle = \lim_{\nu \to 0} \lim_{N \to \infty}
 {\cal P} \left(  R_{e,1,1} Z_{df}  \right),
\end{equation}
and of the auto-correlation function 
$\left\langle R_{e,2}(\tau) R_{e,1}(0) \right\rangle$ given by
\begin{equation}
 {\cal Z}  \left\langle R_{e,2}(\tau) R_{e,1}(0) \right\rangle = \lim_{\nu \to 0} 
 \lim_{N \to \infty}  {\cal P} \left(  R_{e,2,m} R_{e,1,1} Z_{df}  \right) 
\end{equation}
where $\lim\limits_{N \to \infty} \left( \frac{ m\beta}{N}\right) = \tau$.

As an illustration for the evaluation of the temperature Green's function, 
the paper will describe the integration of only one electron propagator, which   
will then simply be noted ${\cal G}$ for compactness. However, the presentation 
of the procedure is quite general, and all the key steps in the derivation of 
the Green's function are expounded. Below, we consider the auto-correlation 
associated with the creation of an electron with spin up on site 1 at imaginary 
time 0, followed by its annihilation on the same site at the imaginary-time 
variable $\tau$. It may be calculated as
\begin{align}
  {\cal Z}  {\cal G} =
  - \lim_{\nu \to 0} \lim_{N \to \infty} {\cal P} \left( 
\left\langle z_{\uparrow,1,m}^{\phantom{\dagger}}
f_{\uparrow,1,m}^{\phantom{\ast}}
f_{\uparrow,1,1}^{\ast}  z_{\uparrow,1,1}^{\star}   \right\rangle_{\!\!f_{\uparrow},
f_{\downarrow},d} \right).
\end{align}
Thus, the computation of the partition function, expectation values 
and dynamical correlation functions bear strong similarities when they are 
expressed as functional integrals.

\section{Integration over $f_{\sigma}$ fields : $N$-time products}\label{sec:int_f}

For the partition function, or a pure correlation of radial boson fields, integrating 
the fermion fields yields
\begin{align}
   Z_f = \big\langle  1 \big\rangle_{f_{\uparrow},f_{\downarrow}} =
  \det \big[S_{\uparrow}\big] \times  \det \big[S_{\downarrow}\big].
\end{align}
Each determinant may be written as the trace of a time-ordered matrix product~: 
\begin{equation}\label{eq:det_S_sigma}
 \det \big[S_{\sigma}\big] = {\rm Tr} \big[  U_{\sigma,N:1} \big].
\end{equation}
The product represents the imaginary-time evolution operator 
 for  $f_{\sigma}$ pseudofermions between the time steps 1 and $N$, and 
 it is defined, for $n_f \ge n_i$, as 
\begin{align}\label{eq:U_sigma}
 \big[ U_{\sigma,n_f:n_i} \big] & = {\cal T}\left\{ \prod_{n=n_i}^{n_f} 
 \big[K_{\sigma,n}\big] \right\}
 \nonumber \\\
 & =  \big[K_{\sigma,n_f}\big] \big[K_{\sigma,n_f-1}\big] \cdots  \big[K_{\sigma,n_i}\big]  
\end{align}
with
\begin{equation} \label{eq:mat_K}
 \big[K_{\sigma,n}\big] = \begin{bmatrix}
                   1 & 0 & 0 & 0 \\ 
                   0 & L_{\sigma,1,n} & T_{\sigma,1,n} & 0\\
                   0 &  T_{\sigma,2,n} & L_{\sigma,2,n} & 0\\ 
                   0 & 0 & 0 &  L_{\sigma,1,n}  L_{\sigma,2,n} 
                  \end{bmatrix}.
\end{equation}
Please note that Eqs.~(\ref{eq:det_S_sigma})--(\ref{eq:U_sigma}) are equivalent to Eq.~(30) 
in Ref.~\cite{RFHOTK07}.
The three  blocks in $[K_{\sigma,n}]$ describe the dynamics of the states 
involving respectively zero, one, and two auxiliary fermions of spin projection $\sigma$. 
The entry $L_{\sigma,i,n}$ is associated with a fermion $f_{\sigma}$ staying still on 
site $i$ at time step $n$, while $T_{\sigma,i,n}$ corresponds to its hopping 
from site $i-1$ onto site $i$. 

The tensor product
\begin{equation}\label{eq:K_n}
 \big[  K_{n} \big] = \big[  K_{\uparrow,n} \big] \otimes \big[  K_{\downarrow,n} \big]
\end{equation}
thus describes the dynamics of the total system of pseudo\-fermions. And its 
partition function may be expanded as a sum over all multi-fermion state evolutions 
along a closed trajectory~:
\begin{align}\label{eq:Z_f}
 Z_f & =  {\rm Tr} \big[  U_{\uparrow,N:1} \big] \times {\rm Tr}
 \big[  U_{\downarrow,N:1} \big] = {\rm Tr} \left( \big[  U_{\uparrow,N:1} \big]
 \otimes \big[  U_{\downarrow,N:1} \big]\right) 
 \nonumber \\
 & = {\rm Tr} \left( {\cal T} \left\{ \prod_{n=1}^N \big[  K_{n} \big] \right\} \right).
\end{align}
The different terms in the trace will be called $N$-time products.
They are products of matrix elements representing the position of
each pseudo\-fermion at every time step.
Note that, according to the structure of the matrices $[K_{\sigma,n}]$, an element 
$L_{\sigma,i,n}$ or $ T_{\sigma,i,n}$ can only be followed by either $L_{\sigma,i,n+1}$ 
or $T_{\sigma,i+1,n+1}$. 
Explicitly, the contributions to $Z_f$ from the sectors with zero to four fermions are 
\begin{align}
 \left\{ \begin{array}{ll} 
 0 : &  1\\ 
 1 : & {\rm Tr} \left[ \begin{smallmatrix} L & T \\ T & L\end{smallmatrix} \right]_{\uparrow}
 + \; {\rm Tr}   \left[ \begin{smallmatrix} L & T \\ T & L\end{smallmatrix} \right]_{\downarrow}\\
 2 : & [LL]_{\uparrow} + [LL]_{\downarrow} 
 + {\rm Tr} \left( \left[ \begin{smallmatrix} L & T \\ T & L\end{smallmatrix} \right]_{\uparrow}
 \otimes \left[ \begin{smallmatrix} L & T \\ T & L\end{smallmatrix} \right]_{\downarrow} \right) \\
 3 : & [LL]_{\uparrow}
 {\rm Tr}  \left[ \begin{smallmatrix} L & T \\ T & L\end{smallmatrix} \right]_{\downarrow}
 + \;[ LL ]_{\downarrow} 
 {\rm Tr}  \left[ \begin{smallmatrix} L & T \\ T & L\end{smallmatrix} \right]_{\uparrow}\\
 4 : & [LL]_{\uparrow} \; [LL]_{\downarrow} \end{array} \right.
\end{align}
 where the time-ordered matrix product
\begin{equation} \label{lt_mat_prod}
 \left[ \begin{smallmatrix} L & T \\ T & L\end{smallmatrix} \right]_{\sigma}  
 \equiv {\cal T} \left\{ \prod_{n=1}^N 
 \begin{bmatrix} L_{\sigma,1,n} & T_{\sigma,1,n} \\ 
 T_{\sigma,2,n} & L_{\sigma,2,n}\end{bmatrix}
 \right\}
 \end{equation}
 yields all the discrete-time evolutions of a single fermion with spin projection $\sigma$,
 while 
\begin{equation} \label{ll_prod}
 [ LL ]_{\sigma} \equiv  \prod_{n=1}^N  L_{\sigma,1,n}  L_{\sigma,2,n}
\end{equation}
is the sole possibility for a pair of them.

As shown in Appendix~\ref{app:det_min}, the unnormalized correlation 
$ \big\langle  z_{\uparrow,1,m}^{\phantom{\star}} f_{\uparrow,1,m}^{\phantom{\ast}}
f_{\uparrow,1,1}^{\ast}  z_{\uparrow,1,1}^{\star} \big\rangle_{\!\! f_{\uparrow}} $
may be computed as the trace of the matrix 
\begin{align}
 & \big[ G_{\uparrow}  \big]
 =   z_{\uparrow,1,m}^{\phantom{\star}} z_{\uparrow,1,1}^{\star} \big[ U_{\uparrow,N:m+1} \big] \;  \big[ F_{1} \big] \;
 \big[ U_{\uparrow,m:2} \big] \; \big[ F_{1}\big]^{\dagger}  \; \big[ K_{\uparrow,1} \big]
\end{align}
where   
\begin{equation}
 \big[ F_{1} \big]^{\dagger} = \begin{bmatrix}
                 \; 0 \quad & \quad 0 \quad & \quad 0 \quad & \quad 0 \; \\
                 \; 1 \quad & \quad 0 \quad & \quad 0 \quad & \quad 0 \; \\
                 \; 0 \quad & \quad 0 \quad & \quad 0 \quad & \quad 0 \; \\
                 \; 0 \quad & \quad 0 \quad & \quad 1 \quad & \quad 0 \;
                   \end{bmatrix}, \quad                  
 \big[ F_{1} \big] = \begin{bmatrix}
                 \; 0 \quad & \quad 1 \quad & \quad 0 \quad & \quad 0 \; \\
                 \; 0 \quad & \quad 0 \quad & \quad 0 \quad & \quad 0 \; \\
                 \; 0 \quad & \quad 0 \quad & \quad 0 \quad & \quad 1 \; \\
                 \; 0 \quad & \quad 0 \quad & \quad 0 \quad & \quad 0 \;
                   \end{bmatrix}
\end{equation}
enact the creation and annihilation, respectively, of a particle 
on site $1$, within the basis $\big\{ |0\rangle,$ $f_{\uparrow,1}^{\dagger}|0\rangle,$
$f_{\uparrow,2}^{\dagger}|0\rangle,$ $f_{\uparrow,1}^{\dagger}f_{\uparrow,2}^{\dagger}
|0\rangle \big\}$. Hence, the full integration over pseudofermion fields yields
\begin{equation}
 G_f = \left\langle z_{\uparrow,1,m}^{\phantom{\star}} f_{\uparrow,1,m}^{\phantom{\dagger}}
 f_{\uparrow,1,1}^{\ast}   z_{\uparrow,1,1}^{\star}
 \right\rangle_{\!\! f_{\uparrow},f_{\downarrow}} =
 {\rm Tr}\left( \big[ G_{\uparrow} \big]
 \otimes  \big[  U_{\downarrow,N:1} \big] \right).
\end{equation}
Similarly to $Z_f$, each $N$-time product here is a product of entries $L_{\sigma,i,n}$ 
or $T_{\sigma,i,n}$ describing the pseudofermion positions during an evolution. 
However, the number of matrix elements is not the same for every time step since an 
extra $f_{\uparrow}$ particle is present between $n=2$ and $n=m$. 
Furthermore, the product also includes the supplemental factor $z_{\uparrow,1,1}^{\star}$ 
associated with the addition of the latter, and the factor 
$z_{\uparrow,1,m}^{\phantom{\star}}$ for its removal. Besides, it always
contains $L_{\uparrow,1,m}$ or $T_{\uparrow,1,m}$, which reflects the 
fact that the particle is on site 1 when it is removed from the cluster. And
conversely, there is not any entry $L_{\uparrow,1,1}$ nor $T_{\uparrow,1,1}$ 
because site 1 has to be free of $f_{\uparrow}$ pseudofermion at the 
beginning.

Below, we coin an $N$-time product as regular if any factor $z_{\sigma,i,n}$ 
within it is companied by the matrix element $L_{\sigma,i,n}$, and any 
$z_{\sigma,i,n}^{\star}$ is followed by $L_{\sigma,i,n+1}$. Since the hopping 
term $T_{\sigma,i,n}$ contains 
$z_{\sigma,i,n}^{\star}z_{\sigma,i-1,n-1}^{\phantom{\star}}$, it has then to be   
immediately preceded by $L_{\sigma,i-1,n-1}$ and succeeded by $L_{\sigma,i,n+1}$.
In other words, a regular product does not possess any clustered factors 
$T_{\sigma,i+1,n+1}T_{\sigma,i,n}$, and it corresponds to an evolution during 
which there is not any successive hoppings of the same pseudo\-fermion  
(note that the products with $T_{\sigma,i,N}$ and $T_{\sigma,i+1,1}$ on both 
extremities are irregular for the trace is invariant under a circular shift).
The motive for the above definition is to ensure  the inverse square roots in 
$z$ factors are eventually valued to unity within physical $N$-time products. 
In this prospect, $R_{e,i,n}$ and $R_{-\sigma,i,n}$ have thus to
vanish in $z_{\sigma,i,n}$ when enforcing the constraints  
Eqs.~\ref{eq:q2_KR}. One can easily check that this requirement is 
fulfilled when an $f_{\sigma}$ particle is on the site $i$ at the 
time step $n$, which is equivalent to state the $N$-time product contains the factor 
$L_{\sigma,i,n}$. 
Besides, $R_{e,1,n+1}=1$ or $R_{-\sigma,i,n+1}=1$ must be satisfied as 
well. However, this is always the case because  $z_{\sigma,i,n}$ is associated 
with the removal of an $f_{\sigma}$ pseudofermion from site $i$, so there is 
not any at the next time step.  As for the factor
$z_{\sigma,i,n}^{\star}$, one needs that $R_{e,i,n+1} = R_{-\sigma,i,n+1}=0$
in physical evolutions, while $R_{e,i,n} =1$ or $R_{-\sigma,i,n}=1$. The former
property is enforced if and only if $L_{\sigma,i,n+1}$ is present, while the latter
is always satisfied since there cannot be more than one $f_{\sigma}$ pseudofermion
on each site, so $z_{\sigma,i,n}^{\star}$ and $L_{\sigma,i,n}$ never appear 
together in the same $N$-time product.

\section{Integration over the complex bosonic $d$ field} \label{sec:int_d}

Akin to the fermionic ones the $d$-boson field couples the time steps through 
its dynamics. It furthermore enters $ Z_f $ and $G_f$ through the factors 
$z_{\sigma,i,n}$ and ${z}^{\star}_{\sigma,i,n}$ which are parts 
of the hopping terms. 
Hence, integrating their $N$-time products requires to evaluate weighted 
integrals of products with an equal number of factors $d_i$ and $d_i^{*}$ 
(the correlations of the other kinds of product vanish). 
Since $d_1$ and $d_2$ are not coupled, one only needs the 
partition function of the 'non-interacting' $d_i$ boson
\begin{equation}\label{eq:zd}
Z_{d_i} = \big\langle 1 \big\rangle_{d_i}
\end{equation}
and its unnormalized $k$-particle correlations 
\begin{align}
G_{d_i}\!\left(m_1,\dotsc,m_k \, | \, n_1,\dotsc,n_k\right)
  = \left\langle \! d_{i,m_1}^{\phantom{\dagger}} \!\!\dotsm d_{i,m_k}^{\phantom{\dagger}} 
 d_{i,n_k}^{*} \!\!\dotsm d_{i,n_1}^{*}  \!\right\rangle_{\!\!d_i} 
\end{align}
where 
\begin{equation}
\big\langle \dots \big\rangle_{d_i}  \equiv
 \int \!\!  \prod_{n=1}^N \frac{\dd d_{i,n}^{\phantom{*}}\; \dd d_{i,n}^{*}}{2\ppi\ii}
 \dots  \;  \ee^{-S_{d_i}}.
\end{equation} 
The first of the three physical constraints~(\ref{eq:q2_KR}) imposes that the $d$ boson
is a hard core particle~: On each site there is at most one $d$ boson at a time. 
Therefore, most of the correlations will be canceled by the integral operator ${\cal P}$ at 
the end of the evaluation, and the only relevant products are the ones where the creation and 
annihilation fields strictly alternate when time ordered. 

The thermal averages are cast here in a form which is suitable for the computation with the 
 operator ${\cal P}$. One obtains
\begin{equation}\label{eq:Zdi}
Z_{d_i} = \det\big[S_{d_i}\big]^{-1} = \left( 1 - \xi_i\right)^{-1} 
= \sum_{D_i=0}^{\infty} \xi_i^{D_i}  
\end{equation}
with
\begin{equation}
 \xi_i =  \ee^{-\beta U}  \prod_{n=1}^N  \ee^{\ii \delta \left(- \tilde{\alpha}_{i,n} 
 + \beta_{\uparrow,i,n} + \beta_{\downarrow,i,n} \right)}.
\end{equation}
According to Wick's theorem~\cite{negele}, multi-particle correlations can be expressed  in 
terms of single-particle ones, which are given by the elements of the inverse matrix
$ \big[S_{d_i}^{-1}\big]$. Explicitly,  
\begin{align}\label{wick_th}
 G_{d_i}\left(m_1,\ldots, n_k\right)
 = \sum_{s\in \mathfrak{S}_k} G_{d_i}^{(s)}\left(m_1,\ldots, n_k\right) 
\end{align}
where the sum runs over all complete contractions (multiplied by $Z_{d_i}$) 
\begin{align}\label{G_di}
 G_{d_i}^{(s)}\left(m_1,\ldots, n_k\right) =  Z_{d_i} 
 \big[S_{d_i}^{-1}\big]_{m_{s(k)}, n_k} \dotsm  \big[S_{d_i}^{-1}\big]_{m_{s(1)}, n_1}
\end{align}
and 
\begin{equation} \label{inv_Sd}
 \big[S_{d_i}^{-1}\big]_{m,n} = \left\{ 
 \begin{array}{lr}
  \displaystyle Z_{d_i} \!\!\!\!\!\! \prod_{q\in [\!\![ n+1 , m ]\!\!]} 
  \!\!\!\!\!\!\!\!
  \ee^{\delta \left( - U  - \ii \tilde{\alpha}_{i,q} + \ii \beta_{\uparrow,i,q} 
  + \ii \beta_{\downarrow,i,q} \right)}  &  {\rm if}\;\; m > n, \\
  Z_{d_i} &   {\rm if}\;\; m = n, \\
  \displaystyle Z_{d_i} \!\!\!\!\!\! \prod_{ q\not\in [\!\![ m+1 , n ]\!\!]} 
  \!\!\!\!\!\!\!\!
  \ee^{\delta \left(- U  - \ii \tilde{\alpha}_{i,q} + \ii \beta_{\uparrow,i,q} 
  +  \ii \beta_{\downarrow,i,q}\right) }   &   {\rm if}\;\; m < n  \phantom{,}
 \end{array}
 \right. 
\end{equation}
with $q\not\in [\!\![ m+1 , n ]\!\!]$ standing for 
$q \in [\!\![ 1 , N]\!\!] \setminus [\!\![ m+1 , n ]\!\!]$.
Using the equality
\begin{align}
 Z_{d_i}^{k+1} & = (1 - \xi_i)^{-(k+1)} = \frac{1}{k!} \frac{d^k}{d \xi_i^k}
 \left( \frac{1}{1- \xi_i} \right) \nonumber \\
 & = \frac{1}{k!} \frac{d^k}{d \xi_i^k} \left( \sum_{D_i=0}^{\infty} \xi_i^{D_i} \right) 
 = \sum_{D_i= 0}^{\infty} \binom{D_i + k}{k} \xi^{D_i},
\end{align}
a complete contraction can then be cast into the generic form
\begin{align} \label{cc_expanded}
& G_{d_i}^{(s)}\left(m_1,\ldots, n_k\right) \nonumber \\ 
& = \sum_{D_i= 0}^{\infty} \binom{D_i + k}{k} 
 \prod_{n=1}^N \ee^{\delta ( D_i + D^{(s)}_n ) \left( -U  - \ii \tilde{\alpha}_{i,n}
 + \ii \beta_{\uparrow,i,n} +  \ii \beta_{\downarrow,i,n}
 \right) }
\end{align}
where $D^{(s)}_n$ is written instead of the more rigorous notation 
$D^{(s)}_n\left(m_1,\dotsc,n_k\right)$ 
in order to shorten expressions. 
By taking $k=0$ and $D^{(s)}_n=0$, the expansion~(\ref{eq:Zdi}) of $Z_{d_i}$ is 
recovered. Hence, any thermal average over the $d_i$ field can be cast as a sum of 
terms of the form~(\ref{cc_expanded}), and the latter will be used when discussing 
general properties of $d$-boson correlations. 

As evidenced by the Coulomb amplitude $U$, the sum $D_i + D^{(s)}_n$ 
corresponds to the number of $d$ bosons on site $i$ at time step $n$ during the system 
evolution associated with that particular complete contraction. 
Note that $D_i$ can take any integer value since the double occupancy is not restricted in 
the enlarged Fock space. However, the Lagrange multipliers promote the sum as the 
number of $d$ bosons in the physical constraints~(\ref{eq:q2_KR}). The latter equalities 
impose that $D_i + D^{(s)}_n \le 1$ so that the series in the generic form~(\ref{cc_expanded}) 
actually stops at $D_i=0$ when the operator ${\cal P}$ is applied on a $k$-particle 
correlation, and it is eventually discarded if $D^{(s)}_n > 1$ at any time step. 

\begin{figure}
  \includegraphics[clip=true, width=0.48\textwidth]
  {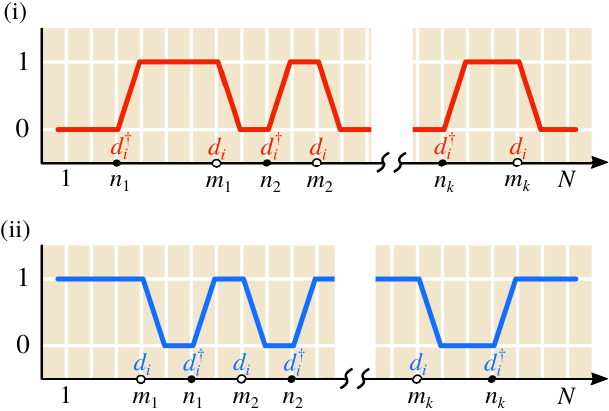}
  \caption{(Color online) Discrete-time variation of the double occupancy $D^{(s)}_n$ in 
  a complete contraction complying with the physical constraints (i) for the time step 
  order   $n_1<\ldots<m_k$, and (ii) for  $m_1<\ldots<n_k$.}
  \label{fig:k-part_corr}
\end{figure}

Considering Eq.~(\ref{inv_Sd}), one thus deduces that in a constraint-compliant complete 
contraction, there cannot be more than one single-particle correlation of the kind 
$\left\langle d_{i,n}^{*} d_{i,m}^{\phantom{\dagger}} \right\rangle$ with $m<n$. In 
this case,  $m$ is the smallest among the $2k$ time steps and $n$ the largest one. 
The other contractions are necessarily of the type $\left\langle d_{i,m}^{\phantom{\dagger}} 
d_{i,n}^{*} \right\rangle$ with $m\ge n$, and their time ranges do not overlap. Hence 
the hard-core nature of the boson imposes that creations and annihilations of $d$ bosons 
must chronologically alternate in a non-vanishing correlation. If there is a
time step $n$ in the sequence at which both $d_i$ and $d_i^{*}$ are evaluated, the 
two corresponding complex variables must be gathered in the same contraction. Indeed, 
such instance can occur when calculating the thermal average of a product containing
a pair $T_{\sigma',i-1,n+1} T_{\sigma,i,n}$. Following the dynamics of pseudo\-fermions
embodied by the matrices $[K_{\sigma,n}]$ (Eq.~(\ref{eq:mat_K})), no factor $L_{\sigma,i,n}$
is then present at time step $n$~: The $f_{\sigma}$ particle does not stand on 
site $i$ at the very instant of its hopping, so the site cannot be doubly occupied. 
Therefore, $D^{(s)}_n$ must vanish in order to comply with physical constraints, which 
is possible only if the variable $d_{i,n}$ is contracted in a equal-time expectation value. 
As a result, in the Wick expansion of a non-vanishing $k$-particle correlation, only one 
complete contraction is relevant since the operator ${\cal P}$ cancels all the other
ones. This one is
$ \prod\limits_{j=1}^k \left[ S_{d_i}^{-1} \right]_{m_j,n_j} $
for  $m_k \ge n_k \ge \ldots \ge m_1 \ge n_1$, and
$\left[ S_{d_i}^{-1} \right]_{m_1,n_k} \prod\limits_{j=1}^{k-1} 
\left[ S_{d_i}^{-1} \right]_{m_{j+1},n_j}$ for $n_k \ge m_k \ge \ldots \ge n_1 \ge m_1$
(see Fig.~\ref{fig:k-part_corr}).

 The remaining details of the $d$-field integration will be discussed in the next 
 sections in conjunction with the filtering by ${\cal P}$ of the joint correlations
\begin{align}
 Z_{df} =\left\langle  Z_f \right \rangle_{\! d} \quad\quad {\rm and} \quad\quad 
 G_{df} =\left\langle  G_f \right \rangle_{\! d} .
\end{align}
 In short, the contributions to $Z_{df}$ from the subspaces with zero to 
 four fermions are
\begin{align}
 \left\{ \begin{array}{ll} 
 0 : & Z_{d_1} Z_{d_2} \\ 
 1 : & \left\langle {\rm Tr}  \left[ \begin{smallmatrix} L & T \\ T & L\end{smallmatrix}
 \right]_{\uparrow} \right\rangle_{\! d} + \;  \left\langle   {\rm Tr}
 \left[ \begin{smallmatrix} L & T \\ T & L\end{smallmatrix} \right]_{\downarrow}
  \right\rangle_{\! d} 
 \\
 2 : & \left([LL]_{\uparrow} + [LL]_{\downarrow} \right)  Z_{d_1} Z_{d_2}
 + \left\langle  {\rm Tr} \left( \left[ \begin{smallmatrix} L & T \\ T & L\end{smallmatrix} \right]_{\uparrow}
 \otimes \left[ \begin{smallmatrix} L & T \\ T & L\end{smallmatrix} \right]_{\downarrow} \right)  \right\rangle_{\! d} \\
 3 : & [LL]_{\uparrow}
  \left\langle {\rm Tr}  \left[ \begin{smallmatrix} L & T \\ T & L\end{smallmatrix}
 \right]_{\downarrow} \right\rangle_{\! d}
 + \;[ LL ]_{\downarrow} 
  \left\langle {\rm Tr}  \left[ \begin{smallmatrix} L & T \\ T & L\end{smallmatrix}
 \right]_{\uparrow} \right\rangle_{\! d} \\
 4 : & [LL]_{\uparrow} \; [LL]_{\downarrow}  Z_{d_1} Z_{d_2} 
 \end{array} \right. .
\end{align}
The terms that contain the product $Z_{d_1} Z_{d_2}$ result from static 
configurations. The other contributions, as well as the computation of $G_{df}$, 
involve higher-order correlations of the $d$ field which are associated with 
fermion hoppings.

\section{Integrations over radial slave-boson fields and constraints} \label{sec:int_P}
The joint partition function $Z_{df}$ and the joint Green's function $G_{df}$ have 
been rewritten as sums over the different $d$-averaged $N$-time products describing 
all the discrete-time evolutions of $f_{\sigma}$ pseudo\-fermions and $d$ bosons within 
the enlarged Fock space. Static contributions are just products of factors $L_{\sigma}$ 
multiplied by $Z_{d_1}Z_{d_2}$, while dynamical $N$-time products contain higher-order 
correlations of $d$ bosons, which are generated when accounting for fermion motion. 
Since any hopping process may be associated with four different sequences of boson number 
changes (see Eq.~(\ref{eq:z_sin})), a $d$-averaged $N$-time product in $Z_{df}$ with 
$K$ hopping factors $T_{\sigma}$ is a linear combination of $4^K$ correlations 
$G_{d_1} G_{d_2}$ ($4^{K+1}$ in the case of $G_{df}$). 
However, among the latter, only one corresponds to the physical variations of site 
occupancy that follow from the evolution of the pseudo\-fermion distribution, and 
as shown below, the operator ${\cal P}$ discards all the other ones.

In order to clarify how physical constraints are eventually enforced within the computation, 
$d_i$-boson correlations are expanded according to Wick's theorem as sums of expressions 
of the form~(\ref{cc_expanded}). 
Lagrange multipliers then enter the complete contractions $G_{d_i}^{(s)}$ 
(and $Z_{d_i}$), as well as the integral operator ${\cal P}$ and the matrix 
elements $L_{\sigma}$. 
As a result, they are gathered in the arguments of imaginary exponentials, which are 
the Fourier transforms of Dirac delta functions (noted $\hat{\delta}$ below) 
implementing Eqs.~(\ref{eq:q2_KR}) on every site and at every time step~\cite{Dao20}. 
For each term $G_{d_1}^{(s_1)} G_{d_2}^{(s_2)}$ of an $N$-time product, 
integrating out Lagrange multipliers yields the product of simple integrals
\begin{align}
& \int_{-\infty}^{\infty} \frac{\delta d\! \alpha_{i,n}}{2\ppi}
 \ee^{\ii \delta \alpha_{i,n} \left(1 - R_{e,i,n} - R_{\uparrow,i,n} - R_{\downarrow,i,n}
 - D_i - D^{(s_i)}_n \right)} \nonumber \\
 & = \hat{\delta} \left(1- R_{e,i,n} - R_{\uparrow,i,n} - R_{\downarrow,i,n} - D_i
 - D^{(s_i)}_n \right)
\end{align}
and
\begin{align}
& \int_{-\infty}^{\infty} \frac{\delta d\! \beta_{\sigma,i,n}}{2\ppi}
 \ee^{\ii \delta \beta_{\sigma,i,n} \left(D_i + D^{(s_i)}_n + R_{-\sigma,i,n} 
 - F_{\sigma,i}(n)\right)} 
 \nonumber \\
 & = \hat{\delta} \left(D_i + D^{(s_i)}_n + R_{\sigma,i,n} - F_{\sigma,i}(n) \right)
\end{align}
where the number of $f_{\sigma}$ fermions $F_{\sigma,i}(n)=1$ if the $N$-time product
contains $L_{\sigma,i,n}$, and $F_{\sigma,i}(n)=0$ otherwise. Because 
radial boson fields are restricted to positive values, the delta functions 
unequivocally bind boson numbers to pseudo\-fermion numbers. 
One thus deduces that when both $L_{\uparrow,i,n}$ and $L_{\downarrow,i,n}$ are 
present, the only finite boson value that is allowed on site $i$ at time step $n$ is 
the double occupancy $D_i + D^{(s_i)}_n=1$. If there is one single factor 
$L_{\sigma,i,n}$, then $R_{\sigma,i,n}=1$. And without any factor $L_{\sigma,i,n}$, 
the constraints imply that only the empty-site amplitude $R_{e,i,n}=1$ is of relevance. 

As a consequence, the full integration of each $N$-time product sets the value of 
double occupancy at every time step, which will thereafter be noted $D_{i}(n)$.
More specifically, for a static $N$-time product, the expansion~(\ref{eq:Zdi}) 
of the partition function $Z_{d_i}$ is actually stripped down to one single term. 
The latter corresponds to the number $D_i =0$ for systems with zero or with 
two pseudo\-fermions of equal spin orientations, and $D_i=1$ in the 
four-pseudo\-fermion case. 
In the dynamical sector, the static evolution of a cluster with one single  
pseudofermion yields $D_i=0$, while $D_i=1$ on the doubly occupied site and 0 
on the other one when the system contains three pseudofermions or two with 
opposite spin projections.
Moreover, for an $N$-time product that contains hopping factors, one can conclude that the 
operator ${\cal P}$ discards all unphysical correlations $G_{d_1}G_{d_2}$. 
Indeed, the delta functions ensure that there is only one boson species at a time 
on each site. So at least one of the two radial-field values, $R_{e,i-1,n}$ 
or $R_{-\sigma,i-1,n}$, entering the numerator of each factor $z_{\sigma,i-1,n-1}$ 
eventually vanishes. Since the same holds true for $z_{\sigma,i,n}^{\star}$ as well, with
$R_{e,i,n}$ or $R_{-\sigma,i,n}$, every factor $T_{\sigma,i,n}$ is then actually 
reduced to one of the four combinations of boson number changes. As a result, the 
final integration over radial fields does not keep more than one of the 
correlations. Furthermore, as discussed in the previous section, the operator 
${\cal P}$ refines the Wick expansion of the remaining one down to its sole complete 
contraction $G_{d_1}^{(s_1)} G_{d_2}^{(s_2)}$ that respects the hard-core property 
of $d$ bosons, with $D_i=0$ and $D_{d_i}^{(s_i)} = D_{i}(n)$ in the 
expansion~(\ref{cc_expanded}). 

At last, since the contribution to the action from the $V$-potential depends only on 
radial boson fields, see Eq.~(\ref{eq:projp2}), the delta functions 
ensure that each $N$-time product is correctly weighted with the physical 
value of the non-local Coulomb interaction energy at every time step.

As shown in Appendix~\ref{app:limit_irr}, the contributions to the thermal averages 
from irregular $N$-time products vanish in the limit $N \to \infty$.
Hence only the results from the full integration of regular ones are of interest, 
and are summarized here. 

We introduce the compact notation 
\begin{equation}
 \tprod{AB\dotsm}{XY\dotsm}_n \equiv A_{\uparrow,n} B_{\uparrow,n} \dotsm 
 X_{\downarrow,n} Y_{\downarrow,n} \dotsm  
\end{equation}
for the product of all the matrix elements that occur at time step $n$ in an $N$-time 
product. The effect of the integration over the $d$ field, followed by the application 
of the integral operator ${\cal P}$, is to replace them according to the following 
mapping -- for the configurations with zero to four pseudo\-fermions~:
\begin{align}\label{eq:mapping}
 0:&  \tprod{1}{1}_n  \mapsto  1 &
 \nonumber \\
 1:&  \tprod{L_i}{1}_n \mapsto  \ee^{-\delta\varepsilon}
   \quad\quad\quad\quad\quad\quad\, \tprod{T_i}{1}_n \mapsto  \delta t_{\nu}
 \nonumber \\
 2:&  \tprod{L_1 L_2}{1}_n \mapsto \ee^{-\delta(2 \varepsilon + V)} 
 \nonumber \\  
 &  \tprod{L_i}{L_i}_n  \mapsto \ee^{-\delta (2 \varepsilon + U)} 
 \quad\quad\quad\;\, \tprod{L_i}{L_{i-1}}_n  \mapsto \ee^{-\delta (2\varepsilon + V)}
 \nonumber \\ 
 &  \tprod{T_i}{L_i}_n \mapsto \delta t_{\nu}\ee^{-\delta \varepsilon} 
 \quad\quad\quad\quad  \tprod{T_i}{L_{i-1}}_n\mapsto \delta t_{\nu}\ee^{-\delta \varepsilon} 
 \nonumber \\ 
 &  \tprod{T_i}{T_i}_n \mapsto  0
 \quad\quad\quad\quad\quad\quad\quad  \tprod{T_i}{T_{i-1}}_n \mapsto  \left(\delta t_{\nu} \right)^2 
 \nonumber \\
 3:&\tprod{L_i}{L_1L_2}_n \mapsto \ee^{-\delta (3 \varepsilon + U + 2 V)}  & 
 \nonumber \\
 &  \tprod{T_i}{L_1L_2}_n \mapsto \delta t_{\nu} \ee^{-\delta (2\varepsilon +V)} & 
 \nonumber \\
 4:&\tprod{L_1L_2}{L_1L_2}_n \mapsto \ee^{-\delta (4 \varepsilon + 2U + 4 V)} & 
\end{align}
with 
\begin{equation}\label{eq:t_nu}
 t_{\nu}=\frac{t}{1+\nu^2}.
\end{equation} 
Because of spin symmetry, the combinations obtained from inverting spin orientations
result into the same values, so they have been omitted. These rules apply  
likewise when computing the partition function ${\cal Z}$ and the 
Green's function ${\cal G}$. However for the latter, there are also the extra 
$z$ factors which are mapped as 
\begin{equation}
z_{\uparrow,1,m}^{\phantom{\star}} z_{\uparrow,1,1}^{\star} \mapsto \left(1+ 
\nu^2\right)^{-1}.
\end{equation}

For the calculation of a correlation $\langle {\cal F} (R_a,R_b,\dots )\rangle$, 
where the function is a product of only radial variables, the procedure 
is only slightly different from the above mapping. Indeed, each $N$-time product 
is now multiplied by the physical value of ${\cal F}$ that follows from the 
evolution of pseudo\-fermions. Since the final value is either 0 or 1, this has 
the effect to discard some of the $N$-time products. This may be implemented by 
nullifying the results of certain combinations from the above list, for the very 
time steps at which the radial fields are evaluated.  
For instance, when  computing the hole density with the factor $R_{e,1,1}$, 
only the mapping for $n=1$ is modified. The affected combinations correspond 
to configurations with $R_{e,1,1}=0$, i.e. with an occupied site 1, which now 
yield a vanishing factor~: 
They are $(L_1\otimes 1)_1$, $(L_1L_2\otimes 1)_1$, $(L_1\otimes L_1)_1$, 
$(L_i\otimes L_{i-1})_1$, $(T_i\otimes L_1)_1$, all the products for three and 
four pseudofermions, and the corresponding ones obtained by flipping spin 
orientations. And for the computation of 
the correlation $ \left\langle R_{e,2,m} R_{e,1,1} \right\rangle$, the list of 
nullified combinations is extended with the ones associated with
an occupied site 2 at $n=m$~: They are $(L_2\otimes 1)_m$, and so on.

\section{Recovering the results derived from the Hamiltonian formulation}
\label{sec:recovering}

At the beginning of the integral evaluation, time steps were intertwined, as
illustrated by the expression of the hopping element $T_{\sigma,i,n}$, see 
Eqs.~(\ref{mat_elem_T}) and (\ref{eq:z_sin}). However, this is no longer the 
case once the computation with the operator ${\cal P}$ is completed. Actually, 
the factors within the end value of every regular $N$-time product do not even 
depend on the time step, apart marginally at $n=1$ and $n=m$ for the correlations 
in which these time steps play a particular role. The projection onto the 
physical subspace of a thermal average may thus be recast into a more compact 
form. 
And as shown below, in the continuous-time limit, the latter results in the 
same expression as the one directly derived from the Hamiltonian in the original 
physical Fock space.  
This demonstrates the validity of the integration procedure within the radial 
gauge formulation.

\subsection{Partition function}
Starting from the pseudo-fermion partition function $Z_f$ given by 
Eq.~(\ref{eq:Z_f}), we have derived the corresponding joint correlation 
$Z_{df}$ and its ${\cal P}$-projection. It contains both regular and irregular 
contributions. However, the latter do not contribute to the partition function 
in the continuous-time limit (see Appendix~\ref{app:limit_irr}), and by neglecting them, 
a simpler expression for the projection of the joint partition function is obtained as
\begin{align}\label{eq:P_of_Zdf}
 {\cal P} (Z_{df}) &={\cal P} \left( \left\langle {\rm Tr} \left( 
 \left[ U_{\uparrow,N:1} \right] \otimes \left[ U_{\downarrow,N:1} \right] 
 \right) \right\rangle_d \right) 
 \nonumber \\
 & =  {\rm Tr} \left( \big[ \kappa \big]^N \right) + o ( 1 ).
\end{align} 
Here $\big[ \kappa \big]$ is the time-independent $16\times 16$ matrix 
that results from applying the mapping Eq.~(\ref{eq:mapping}) on the
infinitesi\-mal-evolution matrix 
$\big[K_{\uparrow,n}\big] \otimes \big[K_{\downarrow,n}\big]$ of the 
pseudo\-fermion system. It reads to the first order in $N^{-1}$, 
\begin{equation}\label{eq:kappa}
 \big[ \kappa \big] = \id{16} - \frac{\beta}{N} \big[ H_{\nu} \big] 
 + o \left( \frac{\id{16}}{N}\right).
\end{equation}
The mapping Eq.~(\ref{eq:mapping}) allows us to work out the matrix elements of 
$\big[ H_{\nu} \big] $. They turn out to be identical to the ones of the matrix 
$\big[ H \big]$ that represents the original Hamiltonian, Eq.~(\ref{eq:hamilton}), 
at the exception of the hopping amplitude which, here, reads $t_{\nu}$ as given 
in Eq.~(\ref{eq:t_nu}), instead of $t$. At this point, the regulator $\nu$ has 
played its role~: It has allowed to handle the irregular contributions by 
rendering them finite, but it has no further impact on the final result when 
the limit $\nu\rightarrow 0$ is taken. One can note that within a representation
that does not include the Kotliar square roots, the exact same calculation would
have resulted in the same expressions with $t$ instead of $t_{\nu}$. 
Explicitly, 
\begin{equation}
 \big[ H_{\nu} \big] = 
 \begin{bmatrix} \big[ H_{\nu}^{(\uparrow:0)} \big] & & & \\
 & \big[ H_{\nu}^{(\uparrow:1a)} \big]  & \big[ H_{\nu}^{(\uparrow:1b)} \big] &  \\
 & \big[ H_{\nu}^{(\uparrow:1c)} \big]  & \big[ H_{\nu}^{(\uparrow:1d)} \big] &  \\
 & & &  \big[ H_{\nu}^{(\uparrow:2)} \big]
 \end{bmatrix}
\end{equation}
where
\begin{align}
 \big[ H_{\nu}^{(\uparrow:0)} \big]
 = \begin{bmatrix}  0 \quad\, & 0 & 0 & 0 \\
 0 \quad\,  &  \varepsilon \; &  - t_{\nu}& 0 \\
0  \quad\, &  -t_{\nu}  & \; \varepsilon & 0 \\
 0 \quad\, & 0 & 0 & 2\varepsilon + V  
 \end{bmatrix}
\end{align}
describes the dynamics of the states without any spin-up electron
-- the reader is reminded that $\varepsilon$ stands for the difference 
in energy between the orbital level and the chemical potential,
\begin{align}
 \big[ H_{\nu}^{(\uparrow:1a)} \big]
 & =  \begin{bmatrix}  \varepsilon  \quad\, & 0 & 0 & 0 \\
 0 \quad\,  &  2\varepsilon +U \; &  -t_{\nu}& 0 \\
0  \quad\,  &  -t_{\nu}  & \; 2\varepsilon + V & 0 \\
 0 \quad\,  & 0 & 0 & 3\varepsilon + U + 2 V  
 \end{bmatrix},
 \nonumber \\
  \big[ H_{\nu}^{(\uparrow:1b)} \big] &= \big[ H_{\nu}^{(\uparrow:1c)} \big] 
  = - t_{\nu} \id{4} ,
 \nonumber \\
 \big[ H_{\nu}^{(\uparrow:1d)} \big] 
 &=  \begin{bmatrix}  \varepsilon  \quad \quad & 0 & 0 & 0 \\
 0 \quad \quad &  2\varepsilon +V \; &  -t_{\nu}& 0 \\
 0 \quad \quad &  -t_{\nu}  & \; 2\varepsilon + U & 0 \\
 0 \quad \quad & 0 & 0 & 3\varepsilon + U + 2 V  
 \end{bmatrix}, 
\end{align}
represent the Hamiltonian of the states with one spin-up electron, and
\begin{align}
 &\big[ H_{\nu}^{(\uparrow:2)} \big] 
 = \begin{bmatrix}  2\varepsilon+V  & 0 & 0 & 0 \\
 0 \quad  &  3\varepsilon +U+2V \; &  -t_{\nu}& 0 \\
 0 \quad  &  -t_{\nu}  & \; 3\varepsilon + U +2V& 0 \\
 0 \quad  & 0 & 0 & 4\varepsilon + 2U + 4 V  
 \end{bmatrix} 
\end{align}
corresponds to the states with two spin-up electrons. 

And using the expansion Eq.~(\ref{eq:kappa}) of the matrix $\big[\kappa\big]$, 
the partition function is finally obtained as 
\begin{align}
\mathcal{Z} & = \lim_{\nu\to 0} \lim_{N\to \infty}
{\mathcal P} \left(  Z_{df} \right)
 = \lim_{\nu\to 0} {\rm Tr}\left(  \lim_{N\to \infty}
 \big[ \kappa \big]^N \right)
 \nonumber \\
& = \lim_{\nu\to 0} {\rm Tr}\left(  \ee^{ -\beta [ H_{\nu} ] } \right)
  = {\rm Tr}\left(  \ee^{ -\beta [ H ] } \right)
\end{align}
with  $\big[ H_{\nu=0} \big] = \big[ H \big]$.

\subsection{Green's function}
As with the evaluation of the partition function, the irregular contributions to 
the Green's function vanish in the continuous-time limit. Hence they can be neglected, 
and using the mapping Eq.~(\ref{eq:mapping}), the projection of the joint Green's 
function $G_{df}$ can be cast as
\begin{align}\label{eq:P_of_Gdf}
  {\cal P}(G_{df}) &= {\cal P} \left( \left\langle {\rm Tr} \left( 
 \left[ G_{\uparrow} \right] \otimes \left[ U_{\downarrow,N:1} \right] 
 \right) \right\rangle_d \right)
 \nonumber \\  
 & = \frac{1}{1+\nu^2}  {\rm Tr} \left( \big[ \kappa \big]^{N-m}
 \big[ F_{\uparrow,1} \big] \big[ \kappa \big]^{m-1} \big[ F_{\uparrow,1} \big]^{\dagger}
 \big[ \kappa \big] \right) + o(1)
\end{align}
where the matrices 
\begin{equation}
 \big[ F_{\uparrow,1} \big]^{\dagger} =  \big[ F_1 \big]^{\dagger} \otimes \id{4},
 \quad\quad
 \big[ F_{\uparrow,1} \big] =  \big[ F_1 \big]\otimes \id{4} 
\end{equation}
represent the creation and the annihilation operators of a spin-up electron on site 1, 
respectively. Accordingly, we finally obtain
\begin{align}
 {\cal Z}{\cal G} & = - \lim_{\nu\to 0} \lim_{N\to \infty} {\cal P} \left( G_{df} \right)
 \nonumber \\
  &= - {\rm Tr} \left( \ee^{ -(\beta -\tau) [ H ] } 
  \big[ F_{\uparrow,1} \big] \ee^{ -\tau [ H ] }  
  \big[ F_{\uparrow,1} \big]^{\dagger} \right),
\end{align}
which is the expected result.

\subsection{Thermal averages of radial slave boson fields~: the example of the radial 
field $R_e$}

The probability of site 1 to be empty at time step 1 follows from
\begin{align}\label{eq:P_of_R1Zdf}
 {\cal P}( R_{e,1,1} Z_{df} ) & = {\cal P} \left( R_{e,1,1} \left\langle {\rm Tr}\left( 
 \left[ U_{\uparrow,N:1} \right] \otimes \left[ U_{\downarrow,N:1} \right] 
 \right) \right\rangle_d \right) 
 \nonumber\\
 & = {\rm Tr}\left( \big[ \kappa \big]^{N-1}
    \big[ \kappa_{e_1} \big] \right) + o(1).
\end{align}
Here $\big[ \kappa_{e_1} \big]$ is the matrix obtained from $\big[ \kappa \big]$
by nullifying the entries that correspond to the infinitesimal evolutions during 
which the site 1 is occupied. 
In the continuous-time limit, it takes the simpler form
\begin{equation}\label{eq:n_e1}
 \big[ n_{e_1} \big]_{i,j} = \delta_{i,1} \delta_{j,1} 
+ \delta_{i,3} \delta_{j,3} + \delta_{i,9} \delta_{j,9} 
+ \delta_{i,11} \delta_{j,11}
\end{equation}
which represents the hole-number operator on site 1. And the hole density on site 1
is then given by
\begin{align}
 {\cal Z} \langle R_{e,1} \rangle 
  &=  \lim_{\nu\to 0} \lim_{N\to \infty} {\cal P} \left( R_{e,1,1} Z_{df} \right)
 \nonumber \\
  &=  {\rm Tr} \left( \ee^{-\beta [H]} \big[ n_{e_1} \big]\right)
\end{align}
which vanishes when the system is fully filled, only. Hence, unlike 
complex bosonic fields, the exact averaged value of a radial slave boson field is 
generically finite, without being in conflict with Elitzur's thereom.

Introducing the matrix $\big[ \kappa_{e_2} \big]$ that is obtained from 
$\big[ \kappa \big]$ by nullifying the matrix elements associated with an 
occupied site 2, the regular part of the projection on the physical subspace 
\begin{align}\label{eq:P_of_R2R1Zdf}
 & {\cal P} \left( R_{e,2,m} R_{e,1,1} Z_{df} \right)
 \nonumber \\
 & = {\cal P} \left( R_{e,2,m} R_{e,1,1} \left\langle {\rm Tr}\left( 
 \left[ U_{\uparrow,N:1} \right] \otimes \left[ U_{\downarrow,N:1} \right] 
 \right) \right\rangle_d \right) 
   \nonumber \\ 
 & = {\rm Tr} \left( \big[ \kappa \big]^{N-m}
   \big[ \kappa_{e_2} \big]  \big[ \kappa \big]^{m-1}
   \big[ \kappa_{e_1}  \big] \right) + o(1).
\end{align}
Since the continuous-time limit of $\big[ \kappa_{e_2} \big]$ is the matrix
\begin{equation}\label{eq:n_e2}
\big[ n_{e_2} \big]_{i,j} = \delta_{i,1} \delta_{j,1} 
+ \delta_{i,2} \delta_{j,2} + \delta_{i,5} \delta_{j,5} 
+ \delta_{i,6} \delta_{j,6}
\end{equation}
of the hole-number operator on site 2, 
\begin{align}
 {\cal Z} \langle R_{e,2}(\tau) R_{e,1}(0) \rangle
 &= \lim_{\nu\to 0} \lim_{N\to \infty} {\cal P} \left( R_{e,2,m} R_{e,1,1} Z_{df} \right)
 \nonumber \\
 &= {\rm Tr} \left( \ee^{-(\beta-\tau) [H]} \big[ n_{e_2} \big]
 \ee^{-\tau [H]} \big[ n_{e_1} \big]\right),
\end{align}
which is the expected result.

Our calculations performed in the radial gauge demonstrate that the Kotliar roots may 
be  translated from their operator forms, Eq.~(\ref{eq:K_roots}), to the corresponding 
regularized expressions in terms of radial slave boson fields, Eq.~(\ref{eq:z_sin}). 
The radial representation allows to overcome the hurdle of the normal ordering procedure 
for the square roots. Thus, we have not only obtained the exact partition function, but 
also the correct Green's function as well as the proper correlation functions. 

\section{Summary and conclusion}\label{sec:conclusion}

Summarizing, we have tested the Kotliar and Ruckenstein slave-boson representation
by thoroughly calculating exactly the functional integral formulation of 
thermodynamical and dynamical properties of the finite-$U$ extended Hubbard model 
on a two-site cluster. Our study is focused on the radial gauge, which is here shown 
to be free of redundant degrees of freedom. 
Our calculations demonstrate that the formulation~(Eq.~(\ref{eq:th_average})), that
has been set up from the outset, is faithful. In particular, it remedies the apparent 
shortcoming of the KR representation related to the normal ordering procedure.
Indeed, accounting for it when the Kotliar roots are included in the calculation 
is a formidable task. Yet, our work shows that it is unnecessary as, starting from 
the original formulation~\cite{KR}, it is possible to rewrite the arguments of the roots 
in terms of radial slave bosons, only, without any loss of generality.
This applies to the evaluation of the partition 
function, averaged values and correlation functions of radial slave-boson fields,
and the electronic Green's function, to quote a few (see Sec.~\ref{sec:recovering}). 
We further obtained that the Kotliar roots need to be properly regularized within 
the discrete-time computation of the functional integrals (see Eq.~(\ref{eq:z_sin}))
to cope with otherwise singular values associated to multiple hopping processes 
occurring at consecutive time steps. We then showed that they are not numerous enough 
to yield a finite contribution to the expectation values in the continuous-time limit. 
Hence,  this representation is well defined. Accordingly, the criticisms raised by 
Sch{\"o}nhammer~\cite{Schonhammer1990} are answered, and, recalling that radial slave 
bosons do not undergo Bose condensation, the numerous works based on the saddle-point 
approximation are put on a firmer ground.
Let us also emphasize that using radial slave bosons largely simplifies the handling 
of non-local interactions as they naturally arise as terms quadratic in the radial 
fields, akin to the Hubbard interaction. 
Furthermore, we do not expect further hurdles to appear when tackling larger clusters. 
One may also perform the same calculations without introducing the Kotliar roots. 
In that case, no unexpected difficulty arises while the irregular contributions are 
not singular any longer, though the complex and dynamical $d$ boson field keeps 
the calculation of the partition function far more complicated than in the $U=\infty$ 
case. In that limit, all bosons are radial fields, which have no dynamics on
their own. Regarding the extended Hubbard model, the present representation paves the way to better controlled
calculations of charge fluctuations in the thermodynamic limit.

\appendix

\section{Pseudofermion  Green's function}
\label{app:det_min}
Standard results for Gaussian integrals over Grassmann variables~\cite{negele} 
yield the correlation
\begin{align}
 \left\langle f_{\uparrow,1,m}^{\phantom{\dagger}} f_{\uparrow,1,1}^{\ast}
 \right\rangle_{\!\! f_{\uparrow}}
 = \big[ S^{-1}_{\uparrow} \big]_{2m-1,1} \det \big[ S_{\uparrow} \big]
\end{align}
which is also equal to the minor ${\cal M}_{1,2m-1}$ of $\big[ S_{\uparrow} \big]$
obtained by removing the first row and the $(2m-1)$-th column. 
In order to show the latter can be computed as 
${\rm Tr}\,\big[ G_{\uparrow} \big]/\big(z_{\uparrow,1,m}^{\phantom{\star}}
z_{\uparrow,1,1}^{\star}\big)$ as well, it is first expressed with the 
help of the middle blocks $\big[ K^{(1)}_{\uparrow,n} \big]$ of the dynamics matrices
$\big[ K_{\uparrow,n} \big]$, see Eq.~(\ref{eq:mat_K}), instead of the blocks
$\big[ M_{\uparrow,n} \big]$.
In the following, the spin subscript $\uparrow$ will be omitted for compactness.
Multiplying by $-1$ its $N$ rows and its $N$ columns that contain the entries $L_{2,n}$, 
the minor can be evaluated as the determinant
\begin{align}
\begin{array}{|ccccccccc|} 
\begin{smallmatrix} 0 & \phantom{0} & 1 \end{smallmatrix} & \phantom{\lrbrac{m}} &
\phantom{\lrbrac{m}} & \phantom{\lrbrac{m}} & \phantom{\lrbrac{m}} &
\phantom{\lrbrac{m}} & \phantom{\lrbrac{m}} & \phantom{\lrbrac{m}} &
\bbrac{K^{(1)}_1 \!}\\
\lrbrac{\,\minus K^{(1)}_2\,} & \id{2} & & & & & & &\\
& \ddots & \ddots & & & & & &\\
& & \ddots & \id{2} & & & & &\\
& & & \lrbrac{\minus K^{(1)}_m} & \begin{smallmatrix} 0 \\[2.5pt] 1 \end{smallmatrix} & & & &\\
& & & & \rbrac{\!\!\minus K^{(1)}_{m+1}\!\!} & \id{2} & & & \\
& & & & &  \lrbrac{\!\!\minus K^{(1)}_{m+2}\!\!} & \ddots & & \\
& & & & & & \ddots & \ddots & \\
& & & & & & & \lrbrac{\minus K^{(1)}_N} & \id{2}
\end{array}
\end{align}
where we have introduced the notations $\tbrac{M}$, $\bbrac{M}$, $\lbrac{M}$, and
$\rbrac{M}$ for the top row, the bottom row, the left column, and the right column 
of a $2\times 2$ matrix $\big[ {M}\big]$, respectively.

The minor is then calculated by repeating the expansions according to the last two 
columns. The first iteration yields
\begin{equation}
 {\cal M}_{1,2m-1} =  {\cal D}_{0,N>} - T_{2,1} {\cal D}_{1,N>} - L_{2,1} {\cal D}_{2,N>}
\end{equation}
where the determinant ${\cal D}_{0,N>}$ is
\begin{align}
 \begin{array}{|ccccccccc|} 
\begin{smallmatrix} 0 & \phantom{0} & 1 \end{smallmatrix} & \phantom{\lrbrac{M}} &
\phantom{\lrbrac{M}} & \phantom{\lrbrac{M}} & \phantom{\lrbrac{M}} &
\phantom{\lrbrac{M}} & \phantom{\lrbrac{M}} & \phantom{\lrbrac{M}} &
\phantom{\lrbrac{M}}\\
\lrbrac{\,\minus K^{(1)}_2\,} & \id{2} & & & & & & &\\
& \ddots & \ddots & & & & & &\\
& & \ddots & \id{2} & & & & &\\
& & & \lrbrac{\minus K^{(1)}_m} & \begin{smallmatrix} 0 \\[2.5pt] 1 \end{smallmatrix} & & & &\\
& & & & \rbrac{\!\!\minus K^{(1)}_{m+1}\!\!} & \id{2} & & & \\
& & & & &  \lrbrac{\!\!\minus K^{(1)}_{m+2}\!\!} & \ddots & & \\
& & & & & & \ddots & \ddots & \\
& & & & & & & \lrbrac{\minus K^{(1)}_{N\minus 1}} & \id{2}
\end{array}
\end{align}
and, for $n>m+1$, we define ${\cal D}_{1,n>}$ as
\begin{align}
\begin{array}{|ccccccccc|} 
\lrbrac{\,\minus K^{(1)}_2\,} & \id{2} &
\phantom{\lrbrac{M}} & \phantom{\lrbrac{M}} & \phantom{\lrbrac{M}} &
\phantom{\lrbrac{M}} & \phantom{\lrbrac{M}} & \phantom{\lrbrac{M}} &
\phantom{\lrbrac{M}}\\
& \ddots & \ddots & & & & & &\\
& & \ddots & \id{2} & & & & &\\
& & & \lrbrac{\minus K^{(1)}_m} & \begin{smallmatrix} 0 \\[2.5pt] 1 \end{smallmatrix} & & & &\\
& & & & \rbrac{\!\!\minus K^{(1)}_{m+1}\!\!} & \id{2} & & & \\
& & & & &  \lrbrac{\!\!\minus K^{(1)}_{m+2}\!\!} & \ddots & & \\
& & & & & & \ddots & \ddots & \\
& & & & & & & \lrbrac{\minus K^{(1)}_{n\minus 1}} & \id{2} \\
& & & & & & & & \tbrac{\minus K^{(1)}_{n}} \
\end{array} 
\end{align}
while ${\cal D}_{2,n>}$ has the same structure as ${\cal D}_{1,n>}$, except for its last 
row where $\tbrac{\minus K^{(1)}_{n}}$ is replaced by $\bbrac{\minus K^{(1)}_{n}}$.

The first term ${\cal D}_{0,N>}$ can be straight away simplified, thanks to the 
blocks $\id{2}$ on the right part of its diagonal. It is equal to the 
determinant ${\cal D}_{1,m<}$, which is defined as
\begin{align}
{\cal D}_{1,n<} =  \begin{array}{|cccc|} 
\begin{smallmatrix} 0 & \phantom{0} & 1 \end{smallmatrix}  & & & \\
\lrbrac{\,\minus K^{(1)}_2\,} & \id{2} & &  \\
&\ddots  & \;\;  \ddots \;\;\;\;    & \\
 & & \lrbrac{\!\!\minus K^{(1)}_{n-1}\!\!}  & \id{2} \\
 & & & \tbrac{\minus K^{(1)}_n}
\end{array} 
\end{align}
for $2\le n < m+1$. The latter follows the recursion relation 
\begin{equation}
 \begin{bmatrix} {\cal D}_{1,n<} \\ {\cal D}_{2,n<} \end{bmatrix} = 
 \begin{bmatrix}  L_{1,n} & & T_{1,n} \\ T_{2,n} & &  L_{2,n} \end{bmatrix}
 \begin{bmatrix} {\cal D}_{1,n-1<} \\ {\cal D}_{2,n-1<} \end{bmatrix}
 = \big[ K^{(1)}_n \big]  \begin{bmatrix} {\cal D}_{1,n-1<} \\ {\cal D}_{2,n-1<} \end{bmatrix}
\end{equation}
with ${\cal D}_{2,n<}$ being identical to ${\cal D}_{1,n<}$, except for its 
last row where $\tbrac{\minus K^{(1)}_{n}}$ is replaced by $\bbrac{\minus K^{(1)}_{n}}$. 
Since 
\begin{equation}
 {\cal D}_{1,2<} = \left| \begin{array}{c} 
\begin{smallmatrix} 0 & \phantom{0} & 1 \end{smallmatrix} \\
 \tbrac{\minus K^{(1)}_2}
\end{array} \right| = L_{1,2}
\quad\;\; {\rm and } \quad\;\; 
 {\cal D}_{2,2<} = \left| \begin{array}{c} 
\begin{smallmatrix} 0 & \phantom{0} & 1 \end{smallmatrix} \\
 \bbrac{\minus K^{(1)}_2}
\end{array} \right| = T_{2,2},
\end{equation}
one then obtains 
\begin{equation}
 {\cal D}_{0,N>} =\begin{bmatrix} 1 & & 0 \end{bmatrix}
 \big[ K^{(1)}_m \big] \, \big[ K^{(1)}_{m-1} \big] \dots \big[ K^{(1)}_{2}\big]
 \begin{bmatrix} 1 \\ 0\end{bmatrix}.
\end{equation}
The block-diagonal structure of the dynamics matrix $\big[ K_n \big]$ allows to  
readily verify that the previous expression can also be written as
\begin{equation}
 {\cal D}_{0,N>} =\big[ E_0\big] \big[ G_{} \big] \big[ E_0\big]^{\dagger}
 / \big(z_{1,m}^{\phantom{\star}} z_{1,1}^{\star}\big)
\end{equation}
where $ \big[ E_0\big] = \big[ \, 1  \;\; 0   \;\; 0 \;\; 0 \,\big] $
is the row vector of the zero-pseudofermion state.

Expanding the determinants ${\cal D}_{1,n>}$ and ${\cal D}_{2,n>}$ according to their last 
two columns yield the same recursion relation as the previously obtained one~:
\begin{equation}
 \begin{bmatrix} {\cal D}_{1,n>} \\ {\cal D}_{2,n>} \end{bmatrix} = 
 \big[ K^{(1)}_n \big]
 \begin{bmatrix} {\cal D}_{1,n-1>} \\ {\cal D}_{2,n-1>} \end{bmatrix}.
\end{equation}
Hence 
\begin{equation}
 \begin{bmatrix} {\cal D}_{1,N>} \\ {\cal D}_{2,N>} \end{bmatrix} = 
  \big[ K^{(1)}_N \big] \, \big[ K^{(1)}_{N-1} \big] \dots \big[ K^{(1)}_{m+2}\big]
 \begin{bmatrix} \minus T_{1,m+1} \\ \minus L_{2,m+1} \end{bmatrix} {\cal D}_{3,m<}
\end{equation}
with
\begin{align}
{\cal D}_{3,m<} =  \begin{array}{|cccc|} 
\mat{\minus K^{(1)}_2} & \id{2} &
\phantom{\lrbrac{M}} & \phantom{\lrbrac{M}} \\
& \ddots & \ddots & \\
& & \ddots & \id{2} \\
& & & \mat{\minus K^{(1)}_m} 
\end{array}  = \prod_{n=2}^m L_{1,n} L_{2,n} 
\end{align}
Here, we have omitted the terms $T_{1,n} T_{2,n}=o\left(\frac{1}{N}\right)$ that
are in the factors $\det \big( \minus \mat{K^{(1)}_n} \big)$~: Their total contribution vanishes 
in the conti\-nuous-time limit. It is then easy to check that 
\begin{align}
 -T_{2,1} {\cal D}_{1,N>} & = \mat{E_1} \, \mat{G} \, \mat{E_1}^{\dagger} 
 / \big(z_{1,m}^{\phantom{\star}} z_{1,1}^{\star}\big),
 \\
 -L_{2,1} {\cal D}_{2,N>} & = \mat{E_2} \, \mat{G} \, \mat{E_2}^{\dagger}
 / \big(z_{1,m}^{\phantom{\star}} z_{1,1}^{\star}\big),
\end{align}
where $ \mat{E_1} = \big[ \, 0 \;\; 1   \;\; 0 \;\; 0 \,\big] $ and 
$ \mat{E_2} = \big[ \, 0  \;\; 0   \;\; 1 \;\; 0 \,\big] $ 
are the row vectors of the states with one pseudofermion on site $1$ and on 
site $2$, respectively.

And since
\begin{equation}
 \mat{E_3} \, \mat{G} \, \mat{E_3}^{\dagger} =0
\end{equation}
with the row vector $ \mat{E_3} = \big[ \, 0  \;\; 0   \;\; 0 \;\; 1 \,\big]$ 
of the state with one pseudofermion on both sites, one can conclude that   
\begin{equation} 
{\cal M}_{1,2m-1} = {\rm Tr} \;\mat{G} 
/ \big(z_{1,m}^{\phantom{\star}} z_{1,1}^{\star}\big).
\end{equation}

\section{Limit of irregular contributions}\label{app:limit_irr}

It is shown that the contribution from irregular products to the matrix 
representation of the evolution operator vanishes in the continuous-time limit. 
This can be understood by reasoning with a fixed number ${k}$ of hopping factors while 
increasing $N$ to infinity. One then finds that the number of irregular products with 
${k}$ hoppings does not grow fast enough to compensate the decreasing factor 
$\delta^k = (\beta /N)^{k}$, so that their sum tends to zero. However, since 
there is an infinite number of ${k}$ values, adding all the vanishingly small 
contributions has to be performed with care. 

First, let us prove the assertion when the infinitesimal evolution operator is 
represented by a $p\times p$ matrix of the form
\begin{equation}\label{eq:mat_K_pzc}
 \mat{\mathpzc{K}} = \mat{\mathpzc{L}} + \frac{1}{N} \mat{\mathpzc{T}}.
\end{equation}
Here $\mat{\mathpzc{L}}$ is a diagonal matrix with its norm 
$\norm{\mat{\mathpzc{L}}} \le 1$, where the infinity norm of a matrix is 
defined as
\begin{equation}
 \norm{ \mat{\mathpzc{M}} } = \max_{i,j} \left| \mat{\mathpzc{M}}_{i,j} \right|.
\end{equation}
The hopping entries are $O(1/N)$ and they are gathered, among others, in 
$\mat{\mathpzc{T}}/N$. The matrix $\mat{\mathpzc{T}}$ is bounded with 
$\norm{\mat{\mathpzc{T}}} \le r/p$, where $r$ is a positive constant. 
This is the case for the evolution matrix $\mat{\kappa}$ given by 
Eq.~(\ref{eq:kappa}). And it is also true for the matrix 
\begin{equation}\label{eq:knu}
\mat{\mathpzc{K}_{\;\nu}} = \id{16} + \frac{1}{N} \mat{\mathpzc{T}_{\;\nu}}
\quad\quad {\rm with} \quad \mat{\mathpzc{T}_{\;\nu}}_{i,j} = 
\left|\frac{2\beta t}{\nu}\right|,
\end{equation}
which will be used below to bound the irregular contributions to the 
thermal averages discussed in the manuscript. 

The matrix of the evolution operator over a time range $\tau\in ]0, \beta]$
is obtained as the limit of $\mat{\mathpzc{K}}^m$ with the exponent $m$ set 
as the integer part of $\tau N/\beta$. Expanding the power $\mat{\mathpzc{K}}^m$ using
expression~(\ref{eq:mat_K_pzc}) yields a sum of $2^m$ different $m$-time products 
of matrices.
Extending the definition introduced for a product of matrix elements, a matrix 
product is coined irregular when there are at least two adjacent factors 
$\mat{\mathpzc{T}}$ (or at both ends of the product), and regular otherwise. 
Noting the contributions to the expansion from regular and irregular terms as  
 ${\rm Reg}\left(\mat{\mathpzc{K}}^m \right)$ and 
${\rm Irr} \left( \mat{\mathpzc{K}}^m \right)$, respectively, one has the obvious
equality
\begin{equation}
 {\rm Irr} \left( \mat{\mathpzc{K}}^m \right)=  \mat{\mathpzc{K}}^m 
 - {\rm Reg} \left( \mat{\mathpzc{K}}^m \right).
\end{equation}
The definition ensures the irregular matrix products contain all the 
irregular $m$-time products of matrix entries. Actually, they also contain a 
part of the regular ones (e.g. regular products with  alternating hoppings 
of different pseudofermions). However, the contribution from the latter to 
the evolution matrix can be neglected since, as it is shown below, 
$\norm{{\rm Irr} \left( \mat{\mathpzc{K}}^m \right)}$ vanishes when $N~\to~\infty$.

The total number of products with ${k}$ factors $\mat{\mathpzc{T}}$ is 
\begin{equation}
 a_{m,{k}} = \left\{ \begin{array}{ll} 
  \binom{m}{{k}} & \quad {\rm if}\;\; 0 \le  {k}\le m \\
 0 &\quad {\rm if}\;\; {k}>m,
 \end{array} \right. 
\end{equation}
and the number of regular ones  is 
\begin{equation}
 b_{m,{k}} = \left\{ \begin{array}{ll} 1 & \quad {\rm if}\;\;  {k}=0 \\ 
 \binom{m-{k}}{{k}} + \binom{m-{k}-1}{{k}-1} & \quad {\rm if}\;\; 1 \le  {k}\le \frac{m}{2} \\
 0 &\quad {\rm if}\;\; {k}>\frac{m}{2}.
 \end{array} \right.
\end{equation}
Indeed, for $1\le {k} \le \frac{m}{2}$, there are $\binom{m-{k}}{{k}}$ ones with 
$\mat{\mathpzc{L}}$ at the right end, and $\binom{m-{k}-1}{{k}-1}$ ones with 
$\mat{\mathpzc{T}}$ at the right end~: The ones with a factor $\mat{\mathpzc{L}}$ on the 
right may be seen as sequences of $m-{k}$ positions to be filled with ${k}$ products 
$\mat{\mathpzc{T}}\mat{\mathpzc{L}}$ and $m-2{k}$ factors $\mat{\mathpzc{L}}$, 
while the ones with a factor $\mat{\mathpzc{T}}$ on the right may be seen as sequences 
composed of ${k}-1$ products $\mat{\mathpzc{T}}\mat{\mathpzc{L}}$ and $m-2{k}$ factors 
$\mat{\mathpzc{L}}$, sandwiched between $\mat{\mathpzc{L}}$ and $\mat{\mathpzc{T}}$. 

In order to prove that ${\rm Irr} \left( \mat{\mathpzc{K}}^m \right)$
vanishes in the continuous-time limit, let us show that for any real number 
$\epsilon >0$, there  is a rank $N_{\epsilon}$ above which 
 $\norm{{\rm Irr} \left( \mat{\mathpzc{K}}^m \right)} \le \epsilon $.

First, since the norm of a product of two $p\times p$ matrices verifies the inequality
$ \norm{\mat{A} \mat{B} } \le p \norm{\mat{A}} \norm{\mat{B}} $
in the general case, and $ \norm{\mat{A} \mat{B} } \le  \norm{\mat{A}} \norm{\mat{B}} $ 
when one of the matrices is diagonal, the norm of a product with $k$ factors 
$\mat{\mathpzc{T}}/N$ and $m-k$ factors $\mat{\mathpzc{L}}$ is smaller than $(r/N)^k$. 
Hence,
\begin{equation}
 \norm{{\rm Irr} \left( \mat{\mathpzc{K}}^m \right)}
  \le  \sum_{k=2}^m
 \left(a_{m,{k}} - b_{m,{k}}\right) \left( \frac{r}{N} \right)^{{k}}.
\end{equation}

Then, for any integer $n\ge 2$, 
\begin{align}
 \norm{{\rm Irr} \left( \mat{\mathpzc{K}}^m \right)}
& \le  \sum_{{k}=2}^{n}
 \left(a_{m,{k}} - b_{m,{k}}\right) \left( \frac{r}{N} \right)^{{k}} + 
 \sum_{{k}=n + 1 }^{\infty}
 a_{m,{k}} \left( \frac{r}{N} \right)^{{k}} 
  \nonumber \\
  & \le  \sum_{{k}=2}^{n}
 \left(a_{m,{k}} - b_{m,{k}}\right) \left( \frac{r}{N} \right)^{{k}} + 
  \sum_{{k}=n+ 1 }^{\infty} \frac{r^{k}}{{k}!} .
\end{align} 
The infinite sum in the last line is the tail of the power series of $\ee^{r}$. 
Thus there exists an integer $n_{\epsilon}$ such that if $n \ge n_{\epsilon}$, 
then the tail is smaller than $\frac{\epsilon}{2}$, and
\begin{equation}
\norm{{\rm Irr} \left( \mat{\mathpzc{K}}^m \right)}
 \le  \sum_{{k}=2}^{n_{\epsilon}} \left(a_{m,{k}} - b_{m,{k}}\right) \left( \frac{r}{N} \right)^{{k}}
 + \frac{\epsilon}{2}.
\end{equation} 

Now, all the integers $k$ on the right-hand side of the inequality are
lower than the fixed value $n_{\epsilon}$, and the sum can be made as small as 
wished by increasing $N$ (and so $m$, as well). 
Indeed, with $m\ge 2 n_{\epsilon}$, all $k$ values are smaller than $m/2$, and
\begin{align}
0\le \left(a_{m,{k}} - b_{m,{k}}\right) \left( \frac{r}{N} \right)^{{k}}  
& \le 
 \left(\binom{m}{{k}} - \binom{m-{k}}{{k}} \right) \left( \frac{r}{N} \right)^{{k}}
\nonumber \\
& \le 
 \frac{m^{\underline{{k}}} - (m-{k})^{\underline{{k}}}}{N^{k}} \times  \frac{r^{k}}{{k}!} 
\end{align} 
where $n^{\underline{k}} \equiv \frac{n!}{(n-k)!}$.
For a fixed integer $k$, $\frac{m^{\underline{{k}}}}{N^{k}}$ and 
$\frac{(m-{k})^{\underline{{k}}}}{N^{k}}$ converge to the same limit $\tau^{k}$ when 
$N\to\infty$, so their difference vanishes. 
As a consequence, there exists an integer $N_{\epsilon} > 2 n_{\epsilon} \beta/\tau$ such 
that if $N \ge N_{\epsilon}$, then for all ${k} \in [\!\![ 2, n_{\epsilon} ]\!\!]$,
\begin{equation}
0\le \left(a_{m,{k}} - b_{m,{k}}\right) \left( \frac{r}{N} \right)^{{k}} \le  
  \frac{\epsilon}{2} \ee^{-r} \times  \frac{r^{k}}{{k}!}, 
\end{equation}
so that  
\begin{equation}
\norm{{\rm Irr} \left( \mat{\mathpzc{K}}^m \right)}
\le \frac{\epsilon}{2} \ee^{-r} \left(  \sum_{{k}=2}^{n_{\epsilon}} 
\frac{r^{k}}{{k}!} \right) + \frac{\epsilon}{2} \le \epsilon. 
\end{equation}
Hence, one can conclude that
\begin{equation}
 \lim_{N\to \infty} \norm{{\rm Irr} \left( \mat{\mathpzc{K}}^m \right)} = 0.
\end{equation}

Now, Eq.~(\ref{eq:P_of_Zdf}) can be justified. First, in an irregular $N$-time product, there 
are consecutive time-steps at which a pseudo\-fermion hops back and forth. 
For this kind of process, when the physical constraints are satisfied, the evaluation 
 by ${\cal P}$ of a $z$ factor, Eq.~(\ref{eq:z_sin}), yields a term with module  
$\left(|\nu| \sqrt{1+\nu^2}\right)^{-1/2}$ or 0. In contrast, it yields 
$\left(1+\nu^2\right)^{-1/2}$ when the hopping process is regular.  
So a hopping term $T_{\sigma,i,n}$ results in a factor which may always be bounded by 
$\beta t /( |\nu|N)$ when $|\nu|<1$. As a result,
the total irregular contribution to ${\cal P}( Z_{df})$ is smaller in module 
than the trace of ${\rm Irr}\left( \mat{\mathpzc{K}_{\;\nu}}^N \right)$, with the 
matrix $ \mat{\mathpzc{K}_{\;\nu}}$ given by Eq.~(\ref{eq:knu}).
Since the irregular parts of $\mat{\mathpzc{K}_{\;\nu}}^N$ and 
of $\mat{\kappa}^N$ vanish in the continuous-time limit, their traces and the 
irregular contribution to ${\cal P}(Z_{df})$ vanish as well, which 
proves Eq.~(\ref{eq:P_of_Zdf}).

Similarly, the total irregular contribution to ${\cal P}(G_{df})$ may 
be dominated in module by the trace of the irregular part of the matrix product 
\begin{align}
 \mat{\mathpzc{G}_{\nu}}  = \frac{1}{|\nu|} 
  \mat{\mathpzc{K}_{\;\nu}}^{N-m}\mat{F_{\uparrow,1}}
 \mat{\mathpzc{K}_{\;\nu}}^{m-1} \mat{F_{\uparrow,1}}^{\dagger}
  \mat{\mathpzc{K}_{\;\nu}} 
\end{align}
which is
\begin{align}
  {\rm Irr} \mat{\mathpzc{G}_{\nu}}  = & \frac{1}{|\nu|} \left(
    {\rm Irr} \left( \mat{\mathpzc{K}_{\;\nu}}^{N-m} \right) \mat{F_{\uparrow,1}}
   \mat{\mathpzc{K}_{\;\nu}}^{m-1}  \mat{F_{\uparrow,1}}^{\dagger}
  \mat{\mathpzc{K}_{\;\nu}} 
 \right. 
  \nonumber \\ 
 & + 
   \mat{\mathpzc{K}_{\;\nu}}^{N-m} \mat{F_{\uparrow,1}}
  {\rm Irr} \left( \mat{\mathpzc{K}_{\;\nu}}^{m-1} \right) \mat{F_{\uparrow,1}}^{\dagger}
  \mat{\mathpzc{K}_{\;\nu}} 
  \nonumber \\
 & + 
  \left.
    {\rm Irr}\left( \mat{\mathpzc{K}_{\;\nu}}^{N-m}\right) \mat{F_{\uparrow,1}}
 {\rm Irr}\left( \mat{\mathpzc{K}_{\;\nu}}^{m-1}\right) \mat{F_{\uparrow,1}}^{\dagger}
  \mat{\mathpzc{K}_{\;\nu}} 
  \right)
 \nonumber \\ 
 & + O \left( \frac{\id{16}}{N} \right).  
\end{align}
Indeed, expanding the product $\mat{\mathpzc{G}_{\nu}}$ by using Eq.~(\ref{eq:knu}) 
(without commuting the matrices nor simplifying the products of identity matrices), 
one finds that the majority of irregular $(N+2)$-time matrix products are the ones 
that contain at least a pair of adjacent factors $\mat{\mathpzc{T}_{\;\nu}}$ either 
on the left or on the right of the factor $\mat{F_{\uparrow,1}}$, or on both sides. 
The remaining irregular ones belong to the set of products with a factor 
$\mat{\mathpzc{T}_{\;\nu}}$ adjacent to $\mat{F_{\uparrow,1}}$ or $\mat{F_{\uparrow,1}}^{\dagger}$, or a pair of $\mat{\mathpzc{T}_{\;\nu}}$ on both ends of the product. 
It is straightforward to show that their sum is $O(N ^{-1})$. For the latter types, 
the sums of the irregular $(N+2)$-time products are~: 
\begin{itemize}[$\bullet$]
 \item $\frac{1}{|\nu|N}  \mat{\mathpzc{K}_{\;\nu}}^{N-m-1}  
 \mat{\mathpzc{T}_{\;\nu}} \mat{F_{\uparrow,1}} \mat{\mathpzc{K}_{\;\nu}}^{m-1} 
 \mat{F_{\uparrow,1}}^{\dagger} \mat{\mathpzc{K}_{\;\nu}}$,
 \item $\frac{1}{|\nu|N}  \mat{\mathpzc{K}_{\;\nu}}^{N-m} \mat{F_{\uparrow,1}} 
  \mat{\mathpzc{T}_{\;\nu}}\mat{\mathpzc{K}_{\;\nu}}^{m-2} 
 \mat{F_{\uparrow,1}}^{\dagger} \mat{\mathpzc{K}_{\;\nu}}$,
  \item $\frac{1}{|\nu|N}  \mat{\mathpzc{K}_{\;\nu}}^{N-m} \mat{F_{\uparrow,1}} 
 \mat{\mathpzc{K}_{\;\nu}}^{m-2} \mat{\mathpzc{T}_{\;\nu}}
 \mat{F_{\uparrow,1}}^{\dagger} \mat{\mathpzc{K}_{\;\nu}}$,
 \item $\frac{1}{|\nu|N}  \mat{\mathpzc{K}_{\;\nu}}^{N-m} \mat{F_{\uparrow,1}} 
 \mat{\mathpzc{K}_{\;\nu}}^{m-1}
 \mat{F_{\uparrow,1}}^{\dagger}  \mat{\mathpzc{T}_{\;\nu}}$,
 \item $\frac{1}{|\nu|N^2}  \mat{\mathpzc{T}_{\;\nu}} 
 \mat{\mathpzc{K}_{\;\nu}}^{N-m-1} \mat{F_{\uparrow,1}} \mat{\mathpzc{K}_{\;\nu}}^{m-1} 
 \mat{F_{\uparrow,1}}^{\dagger}  \mat{\mathpzc{T}_{\;\nu}}$.
\end{itemize}
Since the matrix powers $\mat{\mathpzc{K}_{\;\nu}}^{N-m}$ and 
$\mat{\mathpzc{K}_{\;\nu}}^{m-1}$ 
converge while their irregular parts vanish, one can deduce that the matrix 
${\rm Irr} \mat{\mathpzc{G}_{\nu}}$ vanishes in the continuous-time limit. 
And, using similar arguments, the same conclusion may be drawn for the 
irregular part of the matrix product
\begin{align}
 \mat{{\bf G}}  = \frac{1}{1+ \nu^2}
  \mat{\kappa}^{N-m}\mat{F_{\uparrow,1}}
 \mat{\kappa}^{m-2} \mat{F_{\uparrow,1}}^{\dagger}
  \mat{\kappa}, 
\end{align}
which completes the proof of Eq.~(\ref{eq:P_of_Gdf}). 
The demonstrations of Eq.~(\ref{eq:P_of_R1Zdf}) and Eq.~(\ref{eq:P_of_R2R1Zdf}) 
follow the same line of reasoning without further difficulties.


\begin{thebibliography}{99}

\bibitem{Bed86}J.\,G. Bednorz and K.\,A. M\"uller, 
\textit{Z. Physik B} \textbf{1986}, \textit{64}, 189.

\bibitem{Raveau91} B.~Raveau, C.~Michel, M.~Hervieu, and D.~Groult, 
\textit{Crystal Chemistry of High-T$_c$ Superconducting Copper Oxides}, 
Springer Series in Material Science \textbf{15}, Springer-Verlag Berlin, 
Heidelberg, New York (1991). 

\bibitem{Malozemoff05} A.\,P. Malozemoff, J.~Mannhart, and D.~Scalapino,
  \textit{Phys. Today} \textbf{2005}, \textit{58}, 41.

\bibitem{Helmolt93} R.~von Helmolt, J.~Wecker, B.~Holzapfel, L.~Schultz, and
  K.~Samwer, 
  \textit{Phys. Rev. Lett.} \textbf{1993}, \textit{71}, 2331.

\bibitem{Tomioka95} Y.~Tomioka, A.~Asamitsu, Y.~Moritomo, H.~Kuwahara, and Y.~Tokura, 
\textit{Phys. Rev. Lett.} \textbf{1995}, \textit{74}, 5108.

\bibitem{Raveau95} B.~Raveau, A.~Maignan, and V.~Caignaert, 
\textit{J. Solid State Chem.} \textbf{1995}, \textit{117}, 424.

\bibitem{Maignan95} A.~Maignan, C.~Simon, V.~Caignaert, and B.~Raveau, 
\textit{Solid State Commun.} \textbf{1995}, \textit{96}, 623. 

\bibitem{Kawa97}H.~Kawazoe, H.~Yasakuwa, H.~Hyodo, M.~Kurota, H.~Yanagi, and H.~Hosono,
\textit{Nature} \textbf{1997}, \textit{389}, 939.

\bibitem{Li11} L.~Li, C.~Richter, S.~Paetel, T.~Kopp, J.~Mannhart, and R.C.~Ashoori, 
\textit{Science} \textbf{2011}, \textit{332}, 825.

\bibitem{Ter97} I. Terasaki, Y. Sasago, and K. Uchinokura,
\textit{Phys.~Rev.~B} \textbf{1997}, \textit{56}, R12685. 

\bibitem{Mas00} A. C. Masset, C. Michel, A. Maignan, M. Hervieu, O. Toulemonde, 
 F. Studer, B. Raveau, and  J. Hejtmanek, 
 \textit{Phys.~Rev.~B}  \textbf{2000}, \textit{62}, 166. 

\bibitem{Mat01} I. Matsubara, R. Funahashi, T. Takeuchi, S. Sodeoka,
  T. Shimizu, and K. Ueno, 
  \textit{Appl. Phys. Lett.} \textbf{2001}, \textit{78}, 3627.  

\bibitem{Mic07} M. Miclau, J. Hejtmanek, R. Retoux, K. Knizek, Z. Jirak,
  R. Fr\'esard, A. Maignan, S. H\'ebert, M.~Hervieu, and C. Martin, 
  \textit{Chem. Mater.} \textbf{2007}, \textit{19}, 4243.

\bibitem{Ohta07} H. Ohta, S. Kim, Y. Mune, T. Mizoguchi, K. Nomura, S. Ohta,
  T. Nomura, Y. Nakanishi, Y. Ikuhara, M. Hirano, H. Hosono, and K. Koumoto,
  \textit{Nat. Mater.} \textbf{2007}, \textit{6}, 129.

\bibitem{Wang13} N. Wang, H. J. Chen, H. C. He, W. Norimatsu, M. Kusunoki, and
  K. Koumoto, \textit{Sci. Rep.} \textbf{2013}, \textit{3}, 3449. 

\bibitem{Gui11} E. Guilmeau, M. Poienar, S. Kremer, S. Marinel, S,  H\'ebert,
R. Fr\'esard, and A. Maignan, \textit{Solid State Commun.} \textbf{2011}, 
\textit{151}, 1798.

\bibitem{Pullar2012} R. C. Pullar,
\textit{Prog. Mater. Sci.} \textbf{2012}, \textit{57}, 1191.

\bibitem{Mizushima1980} K. Mizushima, P. C. Jones, P. J. Wiseman, and J. B. Goodenough,
\textit{Mat. Res. Bull.} \textbf{1980}, \textit{15}, 783.

\bibitem{Nagaura1990} T. Nagaura, and K. Tozawa, 
\textit{Prog. Batteries Sol. Cells} \textbf{1990}, \textit{9}, 209.

\bibitem{Arico2005} A. S. Aric\`o, P. Bruce, B. Scrosati, J.-M. Tarascon, and 
W. van Schalkwijk, 
\textit{Nature Mater.} \textbf{2005}, \textit{4}, 366.

\bibitem{Kang2006} K. Kang, Y. S. Meng, J. Br\'eger, C. P. Grey, and G. Ceder,
\textit{Science} \textbf{2006}, \textit{311}, 977.

\bibitem{Poienar2009} M. Poienar, F. Damay, C. Martin, V. Hardy, A. Maignan, and 
G. Andr\'e, 
\textit{Phys. Rev. B} \textbf{2009}, \textit{79}, 014412.

\bibitem{Maignan2018} A. Maignan and C. Martin, 
\textit{Phys. Rev. B} \textbf{2018}, \textit{97}, 161106.

\bibitem{Wang2003} J. Wang, J. B. Neaton, H. Zheng, V. Nagarajan, S. B. Ogale,
B. Liu, D. Viehland, V. Vaithyanathan, D. G. Schlom, U. V. Waghmare, N. A. Spaldin,
K. M. Rabe, M. Wuttig, and R. Ramesh, 
\textit{Science} \textbf{2003}, \textit{299}, 5613.

\bibitem{McW73} D. B. McWhan, A. Menth, J. P. Remeika, W. F. Brinkman, and 
T. M. Rice, 
\textit{Phys. Rev. B} \textbf{1973}, \textit{7}, 1920.

\bibitem{Hel01} K. Held, G. Keller, V. Eyert, D. Vollhardt, and V. I. Anisimov,
\textit{Phys.~Rev.~Lett.} \textbf{2001}, \textit{86}, 5345.

\bibitem{Lim03} P. Limelette, A. Georges, D. J\'erome, P. Wzietek, P. Metcalf,
 and J. M. Honig, 
 \textit{Science} \textbf{2003}, \textit{302}, 89. 

\bibitem{Tokura1993}  Y. Tokura, Y. Taguchi, Y. Okada, Y. Fujishima, T. Arima,
K. Kumagai, and Y. Iye, 
\textit{Phys. Rev. Lett.} \textbf{1993}, \textit{70}, 2126. 
 
\bibitem{Hub63} J. Hubbard, 
\textit{Proc. R. Soc. London A} \textbf{1963}, \textit{276}, 238.

\bibitem{Kan63} J. Kanamori, 
\textit{Prog. Theor. Phys.} \textbf{1963}, \textit{30}, 275.

\bibitem{Gutz63} M. C. Gutzwiller, 
\textit{Phys. Rev. Lett.} \textbf{1963}, \textit{10}, 169.

\bibitem{And87} P. W. Anderson, 
\textit{Science} \textbf{1987}, \textit{235}, 1196.

\bibitem{Deeg1993} M. Deeg, H. Fehske, and H. B\"uttner, 
\textit{Z. Phys. B} \textbf{1993}, \textit{91}, 31.

\bibitem{Aichhorn2004} M. Aichhorn, H.~G. Evertz, W. von der Linden, and M. Potthoff,
\textit{Phys. Rev. B} \textbf{2004}, \textit{70}, 235107.

\bibitem{Davoudi2006} B. Davoudi and A.-M. S. Tremblay,
\textit{Phys. Rev. B} \textbf{2006}, \textit{74}, 035113.

\bibitem{Kagan2011} M.~Y. Kagan, D.~V. Efremov, M.~S. Marienko, and V.~V. Val'kov,
\textit{JETP Lett.} \textbf{2011}, \textit{93}, 725.

\bibitem{Ayral2013} T. Ayral, S. Biermann, and P. Werner,
\textit{Phys. Rev. B} \textbf{2013}, \textit{87}, 125149.

\bibitem{vanLoon2014} E.~G.~C.~P. van Loon, A.~I. Lichtenstein, M.~I. Katsnelson, O. Parcollet, and H. Hafermann,
\textit{Phys. Rev. B} \textbf{2014}, \textit{90}, 235135.

\bibitem{Lhoutellier2015} G. Lhoutellier, R. Fr\'esard, and A. M. Ole\'s,
\textit{Phys. Rev. B} \textbf{2015}, \textit{91}, 224410. 

\bibitem{Steffen2017} K. Steffen, R. Fr\'esard, and T. Kopp, 
\textit{Phys. Rev. B} \textbf{2017}, \textit{95}, 035143.

\bibitem{Terletska2017} H. Terletska, T. Chen, and E. Gull, 
\textit{Phys. Rev. B} \textbf{2017}, \textit{95}, 115149.

\bibitem{Kapcia2017} K.~J. Kapcia, S. Robaszkiewicz, M. Capone, and A. Amaricci,
\textit{Phys. Rev. B} \textbf{2017}, \textit{95}, 125112.

\bibitem{Schuler2018} M. Sch\"uler, E.~G.~C.~P. van Loon, M.~I. Katsnelson, and T.~O. Wehling,
\textit{Phys. Rev. B} \textbf{2018}, \textit{97}, 165135.

\bibitem{Paki2019} J. Paki, H. Terletska, S. Iskakov, and E. Gull, 
\textit{Phys. Rev. B} \textbf{2019}, \textit{99}, 245146.

\bibitem{Terletska2021} H. Terletska, S. Iskakov, T. Maier, and E. Gull,
\textit{Phys. Rev. B} \textbf{2021}, \textit{104}, 085129.

\bibitem{Roig2022} M. Roig, A.~T. R{\o}mer, P.~J. Hirschfeld, B.~M. Andersen,
\textit{Phys. Rev. B} \textbf{2022}, \textit{106}, 214530.

\bibitem{Philoxene2022} L. Philoxene, V. H. Dao, and R. Fr\'esard,
\textit{Phys. Rev. B} \textbf{2022}, \textit{106}, 235131. 

\bibitem{Linner2023} E. Linn\'er, A.~I. Lichtenstein, S. Biermann, and E.~A. Stepanov,
\textit{Phys. Rev. B} \textbf{2023}, \textit{108}, 035143.

\bibitem{Riegler2023} D. Riegler, J. Seufert, E. H. da Silva Neto, 
P. W\"olfle, R. Thomale, and M. Klett, \textbf{2023},
arXiv:2305.08900. 

\bibitem{AucarBoidi2023} N. Aucar Boidi, K. Hallberg, A. Aharony, and O. Entin-Wohlman,
\textbf{2023}, arXiv:2310.00291.

\bibitem{Kundu2023} S. Kundu, and D. S\'en\'echal,
\textbf{2023}, arXiv:2310.16075.


\bibitem{Barnes76}S.\,E. Barnes, 
\textit{J. Phys. F} \textbf{1976}, \textit{6}, 1375.

\bibitem{Barnes77}S.\,E. Barnes, 
\textit{J. Phys. F} \textbf{1977}, \textit{7}, 2637.

\bibitem{KR}G.~Kotliar and A.\,E. Ruckenstein, 
\textit{Phys. Rev. Lett.} \textbf{1986}, \textit{57}, 1362.

\bibitem{FK} R.~Fr\'esard and G.~Kotliar, 
\textit{Phys. Rev. B} \textbf{1997}, \textit{56}, 12909.

\bibitem{LiWH}T.\,C. Li, P.~W\"olfle, and P.\,J. Hirschfeld, 
             \textit{Phys. Rev. B} \textbf{1989}, \textit{40}, 6817.

\bibitem{FW92} R. Fr\'esard and P. W\"olfle,
              \textit{Int. J. of Mod. Phys. B} \textbf{1992}, \textit{6},  685;
                                         \textbf{1992}, \textit{6}, 3087.
\bibitem{Riegler20} D.~Riegler, M.~Klett, T.~Neupert, R.~Thomale, and
  P.~W\"olfle, 
\textit{Phys. Rev. B} \textbf{2020}, \textit{101}, 235137.

\bibitem{Kotliar-Georges} F.~Lechermann, A.~Georges, G.~Kotliar, and
O.~Parcollet, \textit{Phys. Rev. B} \textbf{2007}, \textit{76}, 155102.

\bibitem{Piefke} C.~Piefke and F.~Lechermann,
            \textit{Phys. Rev. B} \textbf{2018}, \textit{97}, 125154.

\bibitem{RN83a} N.~Read, D.\,M. Newns, 
               \textit{J. Phys. C} \textbf{1983}, \textit{16}, L1055.

\bibitem{RN83b} N.~Read, D.\,M. Newns, 
               \textit{J. Phys. C} \textbf{1983}, \textit{16}, 3273.
  
\bibitem{NR87} D.\,M. Newns and N.~Read, 
               \textit{Adv. in Physics} \textbf{1987}, \textit{36}, 799.

\bibitem{RFTK01} R.~Fr\'esard and T.~Kopp, 
                \textit{Nucl. Phys. B} \textbf{2001}, \textit{594}, 769.

\bibitem{RFHOTK07} R.~Fr\'esard, H. Ouerdane, and T. Kopp, 
                   \textit{Nucl. Phys. B} \textbf{2007}, \textit{785}, 286.

\bibitem{Vollhardt1987} D. Vollhardt, P. W\"olfle, and P. W. Anderson, 
\textit{Phys. Rev. B} \textbf{1987}, 35, 6703.

\bibitem{Dao17}  V.~H.~Dao and R.~Fr\'esard, 
                 \textit{Phys. Rev. B} \textbf{2017}, \textit{95}, 165127. 

\bibitem{Mezio2017} A. Mezio, and R. H. McKenzie, 
\textit{Phys. Rev. B} \textbf{2017}, 96, 035121.                
                

\bibitem{Lil90} L. Lilly, A. Muramatsu, and W. Hanke,
                    \textit{Phys. Rev. Lett.} \textbf{1990}, \textit{65}, 1379.

\bibitem{Fre91} R. Fr\'esard, M. Dzierzawa, and P. W\"olfle,
                    \textit{Europhys. Lett.} \textbf{1991}, \textit{15}, 325.

\bibitem{Igo13} P.~A. Igoshev, M.~A. Timirgazin, A.~K. Arzhnikov, and
  V.~Y. Irkhin,  \textit{JETP Lett.} \textbf{2013}, \textit{98}, 150.
  
\bibitem{Fre92} R. Fr\'esard and P. W\"olfle,
                \textit{J. Phys.: Condens. Matter} \textbf{1992}, \textit{4}, 3625.
                    
\bibitem{Doll2} B. M\"oller, K. Doll, and R. Fr\'esard,
                \textit{J. Phys.: Condensed Matter} \textbf{1993}, \textit{5}, 4847.

\bibitem{SeiSi} G. Seibold, E. Sigmund, and V. Hizhnyakov,
                \textit{Phys. Rev. B} \textbf{1998}, \textit{57}, 6937.    

\bibitem{Fle01} M. Fleck, A. I. Lichtenstein, and A.~M. Ole\'s,
                \textit{Phys. Rev. B} \textbf{2001}, \textit{64}, 134528.


\bibitem{Sei02} J. Lorenzana and G. Seibold,
                   \textit{Phys. Rev. Lett.} \textbf{2002}, \textit{89}, 136401;
                                     \textbf{2003}, \textit{90}, 066404;
                                      \textbf{2005}, \textit{94}, 107006.

\bibitem{Rac06a} M. Raczkowski, R.~Fr\'esard, and A.~M. Ole\'s,
                    \textit{Phys. Rev. B} \textbf{2006}, \textit{73}, 174525.


\bibitem{RaEPL} M. Raczkowski, R.~Fr\'esard, and A.~M.~Ole\'s,
                    \textit{Europhys. Lett.} \textbf{2006}, \textit{76}, 128.

\bibitem{Fre02} R. Fr\'esard and M. Lamboley, 
                  \textit{J. Low Temp. Phys.} \textbf{2002}, \textit{126}, 1091.
                        
\bibitem{FW98} R. Fr\'esard and W. Zimmermann, 
                   \textit{Phys. Rev. B} \textbf{1998}, \textit{58}, 15288.

\bibitem{Igo15} P.A. Igoshev, M.A. Timirgazin, V.F. Gilmutdinov,
                A.K. Arzhnikov, V.Yu. Irkhin,
           \textit{J. Phys: Condens. Matter} \textbf{2015}, \textit{27}, 446002.

\bibitem{Ras88} J.~W. Rasul and T. Li,
       \textit{J. Phys. C: Solid State Phys.} \textbf{1988}, \textit{21}, 5119.

                    
\bibitem{Lav90} M. Lavagna,
                \textit{Phys. Rev. B} \textbf{1990}, \textit{41}, 142.
                        
\bibitem{Li94} T. Li and P. B\'enard, 
               \textit{Phys. Rev. B} \textbf{1994}, \textit{50}, 17837.  
   
\bibitem{Jol91} Th. Jolic{\oe}ur and J. C. Le Guillou,
                \textit{Phys. Rev. B} \textbf{1991}, \textit{44}, 2403.
    
                           
\bibitem{Kot92} Y. Bang, C. Castellani, M. Grilli, G. Kotliar, R. Raimondi,
                and Z. Wang,
                \textit{Int. J. of Mod. Phys. B} \textbf{1992}, \textit{6}, 531;
                 Proceedings of the Adriatico Research Conference and
                 Miniworkshop \textit{Strongly Correlated Electrons Systems
                 III}, eds. Yu Lu, G. Baskaran, A. E. Ruckenstein, E.
                 Tossati (World Scientific Publishing Co., Singapore, 1992).   
                            

                 
\bibitem{Dao18}  V.~H.~Dao and R.~Fr\'esard,
                 \textit{Acta Phys. Pol. A} \textbf{2018}, \textit{133}, 336. 
                 
\bibitem{Noatschk2020} K. Noatschk, C. Martens, and G. Seibold,
                       \textit{J. Supercond. Nov. Magn.} \textbf{2020}, \textit{33}, 2389.
  
\bibitem{Perelomov1986} A. Perelomov, 
  \textit{Generalized Coherent States and Their Applications} 
  (Springer Berlin, Heidelberg, 1986).
  
\bibitem{Dao20}  V.~H.~Dao and R.~Fr\'esard, 
                 \textit{Ann. Phys. (Berlin)} \textbf{2020}, \textit{532},
                 1900491.
                 
\bibitem{RFTK12} R.~Fr\'esard and T.~Kopp, 
                 \textit{Ann. Phys. (Berlin)} \textbf{2012}, \textit{524}, 175. 

                 
\bibitem{Schonhammer1990}
 K. Sch{\"o}nhammer, Phys. Rev. B \textbf{1990}, \textit{42}, 2591.    
                   
\bibitem{negele}  J.~W. Negele and H.~Orland, 
 \textit{Quantum Many-Particle Systems} (Westview Press, Boulder, Colorado, 1998).
                  
\bibitem{Metzner1988} W. Metzner and D. Vollhardt, 
                  \textit{Phys. Rev. B} \textbf{1988} \textit{37}, 7382.                    
                  
\bibitem{Metzner1989a} W. Metzner and D. Vollhardt, 
                  \textit{Phys. Rev. Lett} \textbf{1989} \textit{62}, 324. 
                 
\bibitem{Metzner1989b} W. Metzner, 
                  \textit{Z. Phys. B} \textbf{1989} \textit{77}, 253.                  
                 
\bibitem{Bickers86} N. E. Bickers, Ph.D. Thesis, Cornell University, 1986.

\bibitem{Bickers87} N. E. Bickers,  
              \textit{Rev. Mod. Phys.} \textbf{1987}, \textit{59}, 845.
                   
\bibitem{RFKSTK22} R.~Fr\'esard, K.~Steffen, and T.~Kopp,
      \textit{Phys. Rev. B} \textbf{2022}, \textit{105}, 245118. 

   
        
\end{thebibliography}
\end{document}